\documentclass[ALICE,manyauthors]{cernphprep}

\usepackage[comma,square,numbers,sort&compress]{natbib}
\usepackage{hyperref}
\usepackage{lineno}
\usepackage{multirow}
\usepackage{color}
\usepackage{rotating}
\usepackage{graphics}
\usepackage{soul}
\usepackage{bm}
\usepackage{hyperref}

\newcommand{\snn}{\sqrt{s_{\mathrm{NN}}}}
\newcommand{\mpt}{\langle p_{\mathrm{T}}\rangle}

\newcommand{\dndy}{d$N$/d$y$}
\newcommand{\mmass}{MeV/$\it{c}^2$}
\newcommand{\Tsallis} {L\'{e}vy-Tsallis} 

\newcommand {\gmom}   {\mbox{\rm GeV$\kern-0.15em /\kern-0.12em c$}}
\newcommand {\mmom}   {\mbox{\rm MeV$\kern-0.15em /\kern-0.12em c$}}
\newcommand {\Gmass} {\mbox{\rm GeV$\kern-0.15em /\kern-0.12em c^2$}}

\newcommand{\meandNdeta}{\ensuremath{\langle$d$N_{\mathrm{ch}}$/d$\eta_{\mathrm{lab}}\rangle}}
\newcommand{\dndydpt}{${\rm d}^2N/({\rm d}p_{\rm T}{\rm d}y)$}

\newcommand{\PbPb}{\ensuremath{\mbox{Pb--Pb}}}
\newcommand{\pp}{\ensuremath{\mathrm {p\kern-0.05em p }}}
\newcommand{\pPb}{\ensuremath{\mbox{p--Pb}}}

\newcommand{\pt}{\ensuremath{p_{\mathrm{T}}}}

\newcommand{\MeVc}{\ensuremath{\mathrm{MeV}\kern-0.05em/\kern-0.02em c}}
\newcommand{\GeVc}{\ensuremath{\mathrm{GeV}\kern-0.05em/\kern-0.02em c}}
\newcommand{\GeVcSq}{\ensuremath{\mathrm{GeV}\kern-0.05em/\kern-0.02em c^2}}

\newcommand{\jpsi}{\ensuremath{{\rm J}\kern-0.02em/\kern-0.05em\psi}}

\newcommand{\cL}{\textit{c}}

\begin{document}%

%
\begin{titlepage}
\PHyear{2017}
\PHnumber{017}      
\PHdate{26 January}  

%
\title{Production of $\boldsymbol{\Sigma(1385)^{\pm}}$ and $\boldsymbol{\Xi(1530)^{0}}$ \\
 in p--Pb collisions at $\mathbf{\snn = 5.02}$ TeV}
\ShortTitle{$\Sigma(1385)^{\pm}$ and $\Xi(1530)^{0}$ in p--Pb at $\snn = 5.02 $ TeV}   

\Collaboration{ALICE Collaboration\thanks{See Appendix~\ref{app:collab} 
for the list of collaboration members}}
\ShortAuthor{ALICE Collaboration} 

%
%
%
\begin{abstract}
The transverse momentum distributions of the strange and double-strange hyperon resonances 
\break($\Sigma(1385)^{\pm}$, $\Xi(1530)^{0}$) produced in p--Pb collisions at $\snn = 5.02$ TeV 
were measured in the rapidity range $-0.5< y_\mathrm{CMS}<0$ for event classes corresponding 
to different charged-particle multiplicity densities, $\meandNdeta$. The mean transverse momentum values 
are presented as a function of $\meandNdeta$, as well as a function of the particle masses 
and compared with previous results on hyperon production. The integrated yield ratios of excited to 
ground-state hyperons are constant as a function of $\meandNdeta$. The equivalent ratios to pions
exhibit an increase with $\meandNdeta$, depending on their strangeness content.
\end{abstract}
\end{titlepage}
\setcounter{page}{2}



\newpage

\section{Introduction}
\label{sec:intro}

Hadrons containing one or more strange quarks have been studied extensively over past decades 
in connection with the study of quark-gluon plasma \cite{cite:QGP-Rafel82, cite:StrangeHadron2011}. 
Enhanced hyperon yields were observed in heavy-ion collisions 
with respect to those measured in proton-proton (pp) collisions at the same centre-of-mass energy 
\cite{cite:WA97,cite:NA49-Xi,cite:STAR-Multistrange,cite:strangePbPb}. These enhancements were found to be 
consistent with those expected from thermal statistical model calculations using a grand canonical 
ensemble~\cite{cite:PbPb-GSI}. The canonical \cite{cite:Redlich2002,cite:Kraus2009} 
approach is suggested to explain the relatively suppressed multi-strange baryon yields in 
smaller collision systems such as \pp, proton-nucleus (p--Pb) and peripheral heavy-ion collisions~\cite{cite:Xi_pPb}.  

%
Short-lived resonances, such as K$^{*0}$ and $\Sigma(1385)^{\pm}$, can be used in heavy-ion 
collisions to study the hadronic medium between chemical and kinetic freeze-out \cite{cite:freezeout}. 
Chemical and kinetic freeze-out define the points in time, respectively, when hadron abundances and the 
momenta of particles stop changing. Decay products 
of resonances are subject to re-scattering processes and emerge after kinetic decoupling with little memory of the source. 
Regeneration processes, conversely, increase the resonance yield \cite{cite:mechanism}. If re-scattering processes 
are dominant over regeneration processes, the measured yield of resonances is expected to be reduced. Moreover, the longer 
the time between chemical and kinetic freeze-out, the greater the expected reduction. 

Recently, the ALICE collaboration reported results on  K$^{*0}$, $\phi$, $\Xi^-$ and 
$\Omega^-$ in \pp~and \pPb~collisions ~\cite{cite:Kphi_pp, cite:Xi_pPb, cite:KphipPb}
in addition to \PbPb~data~\cite{cite:strangePbPb, cite:KphiPbPb}. The evolution of the mean 
transverse momenta ($\mpt$) of mesons and multi-strange baryons were presented as a function of 
charged-particle multiplicity and particle mass. The observed decrease of the resonance to ground-state ratio 
K$^{*0}$/K$^-$ has been suggested as an indication of re-scattering processes in the hadronic 
medium, as first observed in Pb--Pb collisions~\cite{cite:KphiPbPb}.

This paper reports on the hyperon resonances $\Sigma(1385)^{\pm}$ ($c\tau = 5.48$~fm, $uus$ or $dds$~\cite{cite:PDG}) 
and  $\Xi(1530)^{0}$ ($c\tau = 22$~fm, $uss$~\cite{cite:PDG}), measured in p--Pb collisions at $\snn = 5.02$ TeV.
The corresponding results for pp collisions have been previously published in~\cite{cite:Xi_pp}. The results presented 
in this paper complement the p--Pb results given in~\cite{cite:Xi_pPb, cite:KphipPb}. The measured \pt~spectra, yields and 
mean transverse momenta are presented for different multiplicity classes. Yield ratios of excited to ground-state hyperons are
studied as a function of event multiplicity and compared with model 
predictions~\cite{ cite:PbPb-GSI, cite:pp-THERMUS, cite:DPMJET, cite:pythia8}. Considering the similar lifetimes of
$\Sigma(1385)^{\pm}$ and K$^{*0}$, a decrease of the $\Sigma(1385)^{\pm}/\Lambda$ ratio, consistent with the decrease observed 
for the K$^{*0}$/K$^-$ ratio, is expected for increasing system sizes. Hyperon to pion ratios are also
presented and compared to the results for ground-state hyperons with the same strangeness contents.

In this paper, the short notations $\Sigma^{*\pm}$ and $\Xi^{*0}$ are adopted for $\Sigma(1385)^\pm$
and $\Xi(1530)^0$. Moreover, the notations $\Sigma^{*\pm}$ and $\Xi^{*0}$ include the respective 
anti-particles, namely $\Sigma^{*\pm}$ includes $\Sigma^{*+}$, $\Sigma^{*-}$, and their anti-particles,
while $\Xi^{*0}$ means $\Xi^{*0}$ and $\overline{\Xi^*}^0$, unless otherwise indicated.

\section{Experimental setup and event selection}
\label{sec:exp}

A description of the ALICE detector and of its performance during the LHC Run 1 (2010--2013)
can be found in \cite{cite:ALICE, cite:ALICEPerformance}. The data sample analysed in this paper 
was recorded during the LHC p--Pb run at $\snn$ = 5.02 TeV in 2013. 
Due to the asymmetric energies of the proton (4 TeV) and lead ion \break(1.57 A TeV) beams, the centre-of-mass 
system in the nucleon-nucleon frame is shifted in rapidity by $\Delta y_{\mathrm{NN}}$ = 0.465 towards the 
direction of the proton beam with respect to the laboratory frame of the ALICE detector \cite{cite:KphipPb}. 
For the analysed p--Pb data set, the direction of the proton beam was towards the ALICE muon spectrometer,
the so-called ``C'' side, standing for negative rapidities; conversely, the Pb beam circulated towards 
positive rapidities, labelled as ``A'' side in the following. The analysis in this paper was carried out at midrapidity, 
in the rapidity window $-0.5 < y_{\mathrm{CMS}} <$ 0.

The minimum-bias trigger during the p--Pb run was configured to select events by requiring a logical OR 
of signals in V0A and V0C \cite{cite:ALICEPerformance}, two arrays of 32 scintillator detectors 
covering the full azimuthal angle in the pseudorapidity regions 2.8 $< \eta_{\mathrm{lab}} <$ 5.1 and 
$-3.7 < \eta_{\mathrm{lab}} < -1.7$, respectively~\cite{cite:rapidity}. In the data analysis it was required to 
have a coincidence of signals in both V0A and V0C in order to reduce the contamination 
from single-diffractive and electromagnetic interactions. This left only Non-Single Diffractive (NSD) events, 
which amount for a total of 100 million events, in the Minimum-Bias (MB) sample corresponding 
to an integrated luminosity of about 50 $\mu$b$^{-1}$.

The combined V0A and V0C information discriminates beam-beam interactions from background collisions in the 
interaction region. Further background suppression was applied in the offline analysis using time information from two 
neutron Zero Degree Calorimeters (ZDC) \cite{cite:ALICEPerformance}, as in previous p--Pb 
analyses~\cite{cite:lambda_pPb}. Pile-up events due to more than one collision in the region of beam interaction were 
excluded by using the Silicon Pixel Detector (SPD) in the Inner Tracking System (ITS)~\cite{cite:ALICEPerformance}. 
The Primary Vertex (PV) is determined by tracks reconstructed in the ITS and Time Projection Chamber (TPC), 
and track segments in the SPD~\cite{cite:ALICEPerformance,cite:rapidity}. MB events are selected when 
the PV is positioned along the beam axis within $\pm$10 cm from the centre of the ALICE detector.

The MB events were divided into several multiplicity classes according to the accumulated charge in the forward 
V0A detector~\cite{cite:centrality-pPb}. The $\Sigma^{*\pm}$ resonances are reconstructed 
in the multiplicity classes \break0-20\%, 20-60\%, and 60-100\%, whereas the $\Xi^{*0}$ 
analysis is carried out in four classes, namely 0-20\%, \break{20-40\%}, 40-60\% and 60-100\%. To each multiplicity class 
corresponds a mean charged-particle multiplicity ($\meandNdeta$), measured at midrapidity 
($|\eta_{\mathrm{lab}}|< 0.5$), as shown in Table \ref{tab:v0estimator}.

\begin{table}[h]
	\centering
	\begin{tabular}{ll}
	\hline
	V0A percentile (\%) &  $\meandNdeta$ \\
	\hline \noalign{\smallskip}
	0-20    &     35.6  $\pm$ 0.8 \\
	20-40  &     23.2  $\pm$ 0.5 \\
	20-60  &     19.7  $\pm$ 0.5 \\
	40-60  &     16.1  $\pm$ 0.4 \\
	60-100&      7.1   $\pm$ 0.2 \\
	0-100  &     17.4 $\pm$ 0.7 \\
	\hline\noalign{\smallskip}
	\noalign{\smallskip}
	\end{tabular}
	\caption{Mean charged-particle multiplicity densities (\meandNdeta) measured at midrapidity 
	($ |\eta_{\mathrm{lab}}|< 0.5$) \cite{cite:rapidity}, corresponding to the 
	multiplicity classes defined using the V0A detector~\cite{cite:centrality-pPb} 
	in \pPb~collisions at $\snn$~=~5.02~TeV.} 
	\label{tab:v0estimator}    
\end{table}

\section{Data analysis}
\label{sec:analysis}
\subsection{Track and topological selections}
\label{subsec:selections}
Table~\ref{tab:PDG} summarizes the relevant information on the measured hyperon resonances, 
namely the decay modes used in this analysis and their branching ratios. In the case of $\Sigma^{*\pm}$, 
all states $\Sigma^{*+}$, $\Sigma^{*-}$, $\overline{\Sigma}^{*-}$ and 
$\overline{\Sigma}^{*+}$ were separately analysed, while the $\Xi^{*0}$ analysis always includes 
the charge-conjugated anti-particle, $\overline{\Xi}^{*0}$ due to the limited statistics of the dataset. 
\begin{table}[h!]
\centering
\resizebox{\textwidth}{!}{   
\begin{tabular}{l|l|l|l|l}
\hline\noalign{\smallskip}
 & Mass (\mmass) & Width (\mmass)  & Decay modes used & Total B.R. (\%) \\
\hline\noalign{\smallskip}
$\Sigma$(1385)$^{+}$ & 1382.80 $\pm$ 0.35 & 36.0 $\pm$  0.7 & $\Lambda \pi^{+}\rightarrow(p\pi^-)\pi^+$ & \multirow{2}{*}{55.6~$\pm$~1.1}\\
$\Sigma$(1385)$^{-}$ &   1387.2 $\pm$ 0.5   & 39.4 $\pm$  2.1 &$\Lambda \pi^{-}\rightarrow(p\pi^-)\pi^-$ &   \\
\hline\noalign{\smallskip}
$\Xi$(1530)$^{0}$ & 1531.80 $\pm$ 0.32 & 9.1 $\pm$ 0.5  &  $\Xi^{-} \pi^{+}\rightarrow(\Lambda\pi^-)\pi^+\rightarrow((p\pi^-)\pi^-)\pi^+$  & 42.6~$\pm$~0.3 \\ 
\hline\noalign{\smallskip}
\noalign{\smallskip}
\end{tabular}
}  
\caption{Properties of the measured resonances and decay modes used in this analysis with total 
branching ratios~\cite{cite:PDG}, obtained as the products of respective branching ratios of daughter particles.}
\label{tab:PDG}    
\end{table}

In comparison with the $\Sigma^{*\pm}$ and $\Xi^{*0}$ analysis carried out in pp collisions 
at $\sqrt{s}$ = 7 TeV~\cite{cite:Xi_pp}, track and topological selections were revised 
and adapted to the p--Pb dataset; this is notably the case for $\Xi^{*0}$.
Pions from strong decays of both $\Sigma^{*\pm}$ and $\Xi^{*0}$ were selected
according to the criteria for primary tracks. As summarized in Table~\ref{tab:primary_selections}, 
all charged tracks were selected with $\pt$~$>$~0.15~ \gmom~ and $|\eta_{\mathrm{lab}}| <0.8$, as 
described in Ref.~\cite{cite:ALICEPerformance}. The primary tracks were chosen with the Distance of 
Closest Approach (DCA) to PV of less than 2 cm along the longitudinal direction (DCA$_z$)  and lower 
than 7$\sigma_r$ in the transverse plane (DCA$_r$), where $\sigma_r$ is the resolution of DCA$_r$. The 
$\sigma_r$ is strongly $\pt$-dependent and lower than 100 $\mu$m for 
$\pt >$~0.5~\gmom~\cite{cite:ALICEPerformance}. To ensure a good track reconstruction quality, 
candidate tracks were required to have at least one hit in one of the two innermost layers (SPD) of 
the ITS and to have at least 70 reconstructed points in the TPC, out of a 
maximum of 159. The Particle IDentification (PID) criteria for all decay daughters are based on the requirement 
that the specific energy loss (d$E$/d$x$) is measured in the TPC within three standard deviations 
($\sigma_\mathrm{TPC}$) from the expected value (d$E$/d$x_{\rm{exp}}$), computed using a Bethe-Bloch 
parametrization~\cite{cite:ALICEPerformance}.
\begin{table}[h!]
\centering
\begin{tabular}{lll}
\hline
Common track &  $|\eta_{\mathrm{lab}}|$ & $<0.8$ \\
 selections & $\pt$ & $> 0.15$ GeV/$c$ \\
& PID $|$(d$E/$d$x)-$(d$E/$d$x)_{\rm{exp}}|$ & $<3$~$\sigma_{\rm{TPC}}$ \\
\hline \noalign{\smallskip}
Primary track& DCA$_z$ to PV         & $<2$ cm \\
 selections & DCA$_r$ to PV         & $<7\sigma_r$ ($p_\mathrm{T}$) \\
& number of SPD points & $\geq 1$ \\
& number of TPC points & $>70$ \\
\hline\noalign{\smallskip}
\noalign{\smallskip}
\end{tabular}
\caption{Track selections common to all decay daughters and primary track selections applied 
to the charged pions from decays of $\Sigma^{*\pm}$ and $\Xi^{*0}$.}  
\label{tab:primary_selections}       
\end{table}

\begin{figure}[h!]
  \subfigure{
    \resizebox{0.05\textwidth}{!}{
}}
  \subfigure{
     \resizebox{0.9\textwidth}{!}{
       \includegraphics[width=15cm]{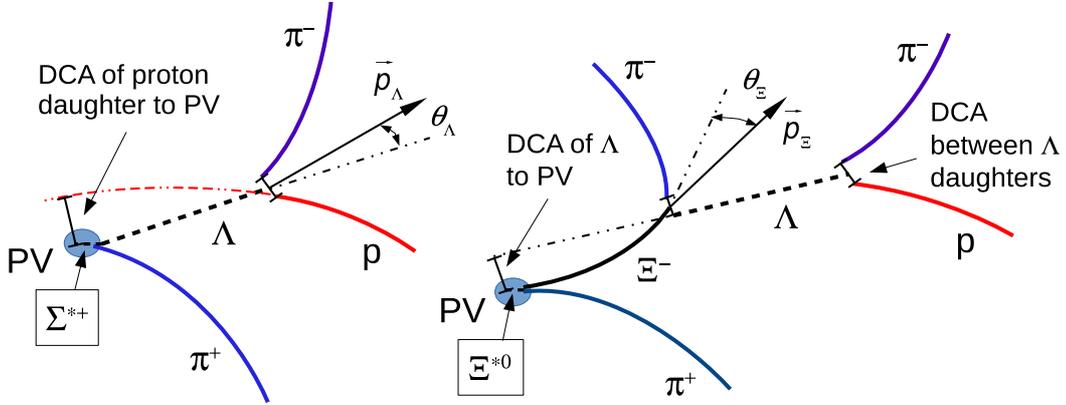}
}
}
   \caption{Sketch of the decay modes for $\Sigma^{*+}$ (left) and $\Xi^{*0}$ (right) and 
   depiction of the track and topological selection criteria.}
  \label{fig:decay}
\end{figure}

Since pions and protons from weak decay of $\Lambda$ ($ \cL \tau = 7.89$ cm~\cite{cite:PDG}) and 
pions from weak decay of $\Xi^{-}$ ($ \cL \tau = 4.91$ cm~\cite{cite:PDG}) are produced away from the PV, 
specific topological and track selection criteria, as summarized in Table~\ref{tab:selections}, were 
applied~\cite{ cite:Xi_pPb,cite:Xi_pp,cite:lambda_pp}.
%
\begin{table}[h!]
\centering
\begin{tabular}{lll}
\hline\noalign{\smallskip}
& $\Sigma^{*\pm}$ & $\Xi^{*0}$ \\
\hline\noalign{\smallskip}
DCA$_r$ of $\Lambda$ decay products to PV  &  $>0.05$ cm & $>0.06$ cm\\
DCA between $\Lambda$ decay products   & $<1.6$~cm & $<1.4$ cm\\
DCA of $\Lambda$ to PV                  & $<0.3$ cm & $>0.015$  cm\\
cos$\theta_\Lambda$      & $>0.99$ & $>0.875$  \\
$r(\Lambda)$      &   1.4 $<r(\Lambda)<$ 100 cm          & 0.2 $<r(\Lambda)<$ 100 cm      \\
$|M_{p\pi} - m_\Lambda|$        &  $<$ 10 \mmass & $<$ 7 \mmass  \\
DCA$_r$ of pion (from $\Xi^{-}$) to PV     && $>0.015$ cm \\
DCA between $\Xi^{-}$ decay products  && $<1.9$ cm\\
cos$\theta_\Xi$    & & $>0.981$    \\
$r(\Xi^-)$     &         & 0.2 $<r(\Xi^-)<$ 100 cm         \\
$|M_{\Lambda\pi} - m_\Xi|$        &&  $<$ 7 \mmass     \\
\hline\noalign{\smallskip}
\noalign{\smallskip}
\end{tabular}
\caption{Topological and track selection criteria.}
\label{tab:selections}. 
\end{table}
%

In the analysis of $\Sigma^{*\pm}$, secondary $\pi$ and p from $\Lambda$ decays were 
selected with a DCA between the two tracks of less than 1.6~cm and 
with a DCA$_r$ to the PV greater than 0.05~cm, to remove most primary tracks. 
For $\Sigma^{*-}$ and $\overline{\Sigma}^{*+}$, the DCA of $\Lambda$ to the PV must be smaller 
than 0.3 cm in order to remove most of the primary weakly-decaying $\Xi(1321)^{-}$ and 
$\overline{\Xi}(1321)^{+}$, which share the same decay channel.
The $\Lambda$ invariant mass ($M_{p\pi}$) was selected within 
$\pm$~10~\mmass~of the Particle Data Group (PDG) value 
($m_\Lambda=1115.683\pm0.006$~\mmass)~\cite{cite:PDG}, the cosine of the pointing 
angle $\theta_\Lambda$ (the angle between the sum of daughter momenta and the line that connects
 the PV and the decay vertex, as shown in Fig.~\ref{fig:decay}) was requested to be greater than 0.99, 
and the radius of the fiducial volume $r(\Lambda)$ (the distance between the PV and the decay vertex) 
was requested to be between 1.4 and 100 cm.

In the analysis of $\Xi^{*0}$, $\Lambda$ and $\pi$ from $\Xi^{-}$ were 
selected with a DCA of less than 1.9~cm and with a DCA$_r$ to the PV greater than 0.015 cm. 
The $\Lambda$ daughter particles ($\pi$ and p) were required to have a DCA$_r$ to the PV greater 
than 0.06 cm, while the DCA between the two particles was required to be less than 1.4 cm. Cuts on 
the invariant mass, the cosine of the pointing angle ($\theta_\Lambda$, $\theta_\Xi$) and the radius of 
the fiducial volume ($r(\Lambda)$, $r(\Xi)$) in Table~\ref{tab:selections} were applied to optimize 
the balance of purity and efficiency of each particle sample. 

\subsection{Signal extraction}
\label{subsec:signal}
\begin{figure*}[tb!]
 \begin{minipage}[c]{\linewidth}
  \subfigure{
    \resizebox{0.5\textwidth}{!}{
      \includegraphics{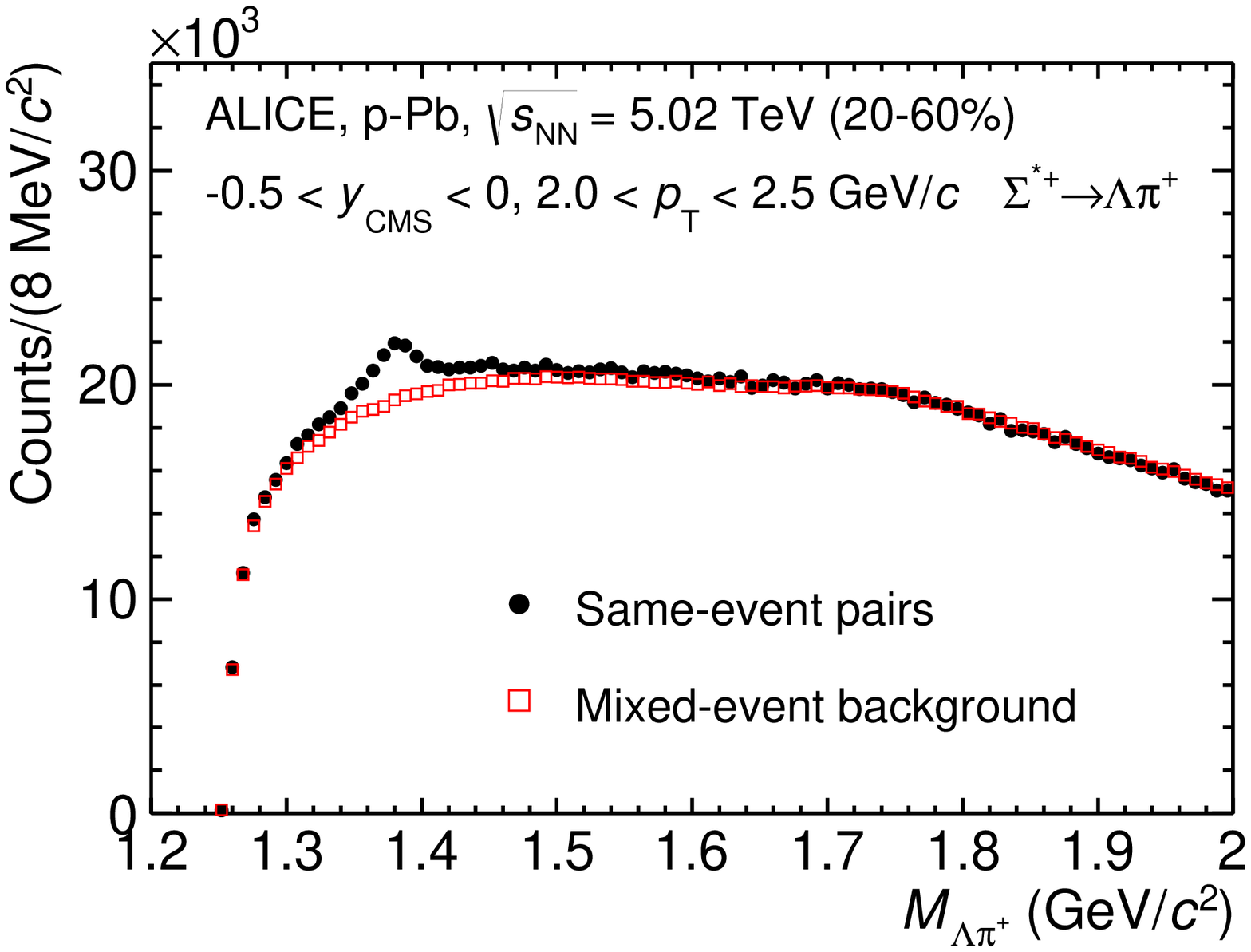}
    }
  } 
  \subfigure {
    \resizebox{0.5\textwidth}{!}{
      \includegraphics{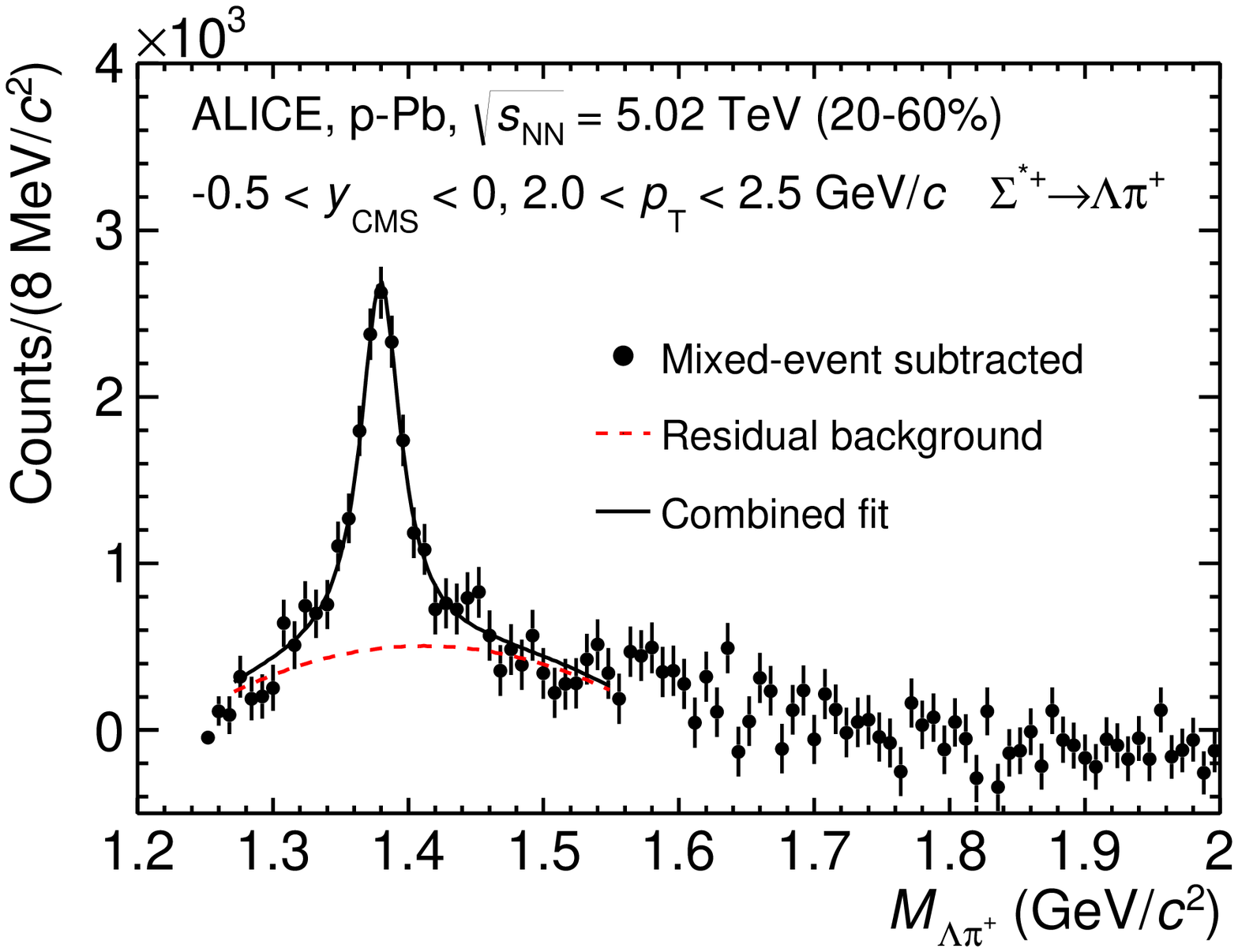}
    }
  }   
\caption{(Left) The $\Lambda\pi^{+}$~invariant mass distribution (Same-event pairs) in 
2.0~$<$~\pt~$<$~2.5~\gmom~and for the multiplicity class 20-60\%. The background shape, 
using pairs from different events (Mixed-event background), 
is normalised to the counts in 1.9~$<M_{\Lambda\pi}<$~2.0~\Gmass. (Right) The  invariant mass 
distribution after subtraction of the mixed-event background. The solid curve 
represents the combined fit, while the dashed line describes the residual background.
}
  \label{fig:invmassSigmaStarPlus}
\end{minipage}
\end{figure*}

\begin{figure*}[tb!]
 \begin{minipage}[c]{\linewidth}
  \subfigure{
    \resizebox{0.5\textwidth}{!}{
      \includegraphics{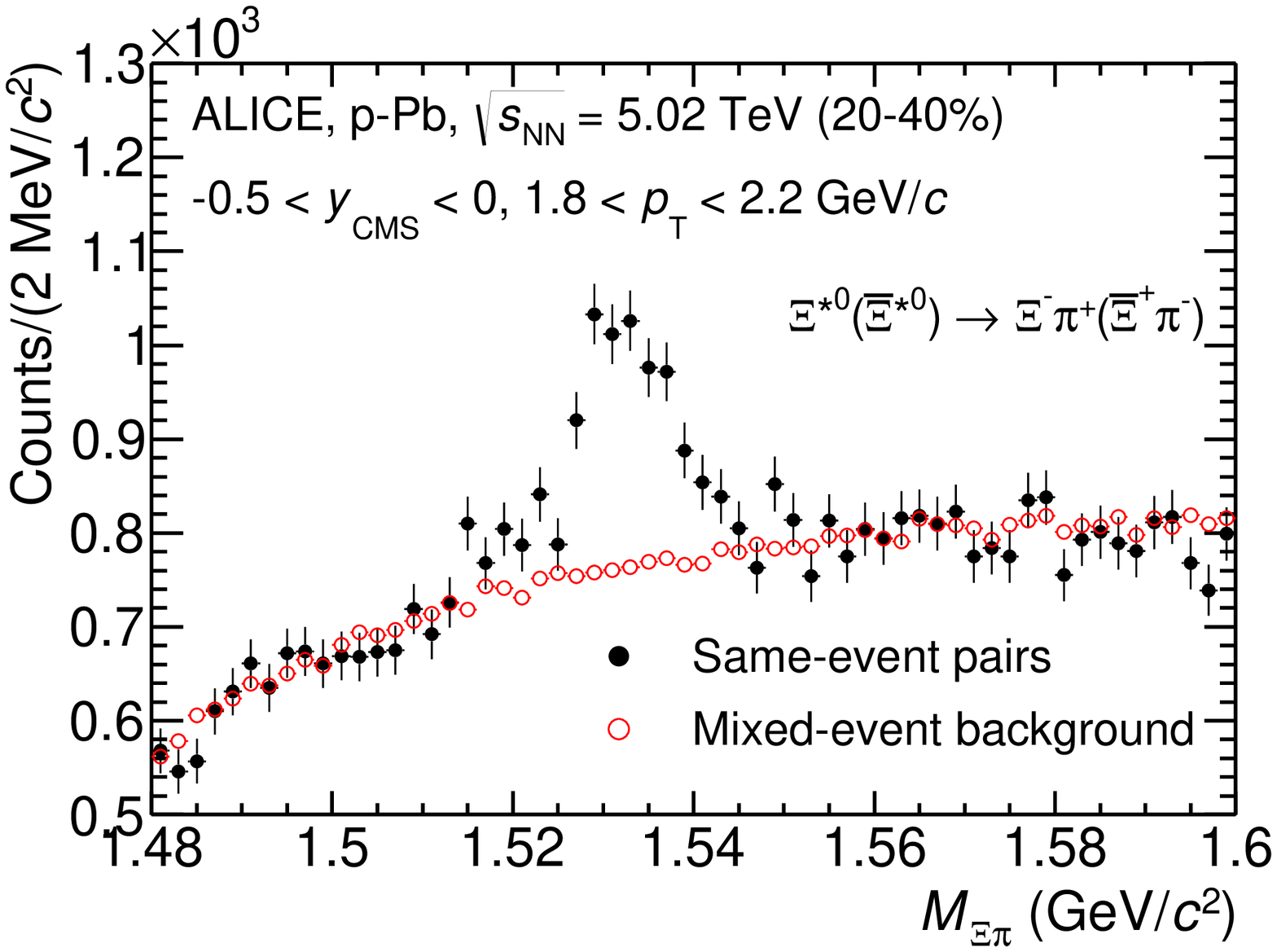}
    }
  } 
  \subfigure {
    \resizebox{0.5\textwidth}{!}{
      \includegraphics{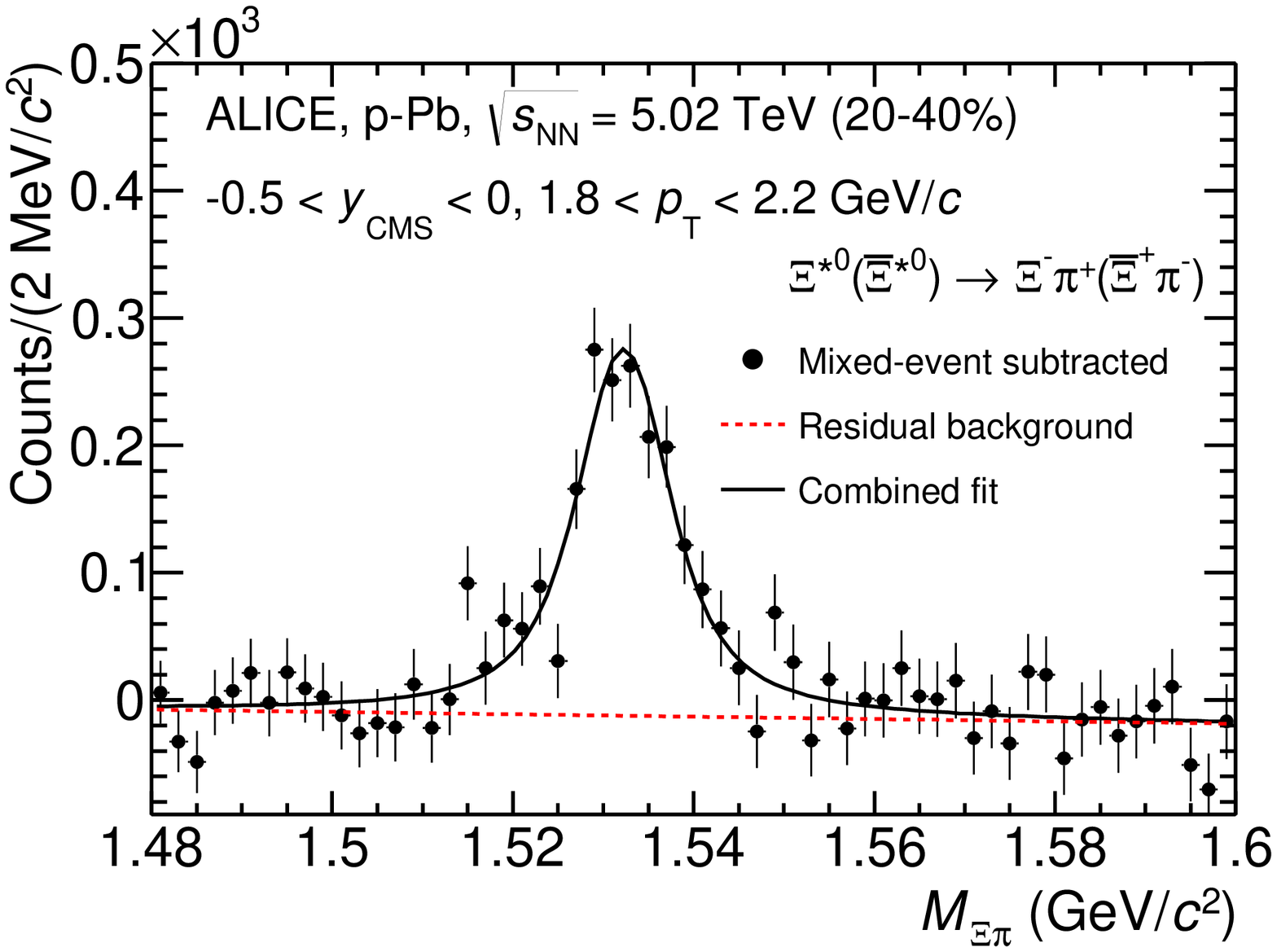}
    }
  }   
\caption{(Left) The $\Xi^{\mp}\pi^{\pm}$ invariant mass distribution (Same-event pairs) in 
1.8~$<$~\pt~$<$~2.2~\gmom~and for the multiplicity class 20-40\%. The background shape, 
using pairs from different events (Mixed-event background), is normalised to the counts in 
1.49~$<$~$M_{\Xi\pi}$~$<$~1.51~\Gmass~and 1.56~$<$~$M_{\Xi\pi}$~$<$~1.58~\Gmass. 
(Right) The invariant mass distribution after subtraction of the mixed-event background. 
The solid curve represents the combined fit, while the dashed line describes the residual background.}
  \label{fig:invmassXiStar}
\end{minipage}
\end{figure*}

The $\Sigma^{*\pm}$ and $\Xi^{*0}$ signals were reconstructed by invariant-mass analysis 
of candidates for the decay products in each transverse momentum interval of the resonance 
particle, and for each multiplicity class. Examples of invariant-mass distributions are presented in 
the left panels of Figs.~\ref{fig:invmassSigmaStarPlus} and~\ref{fig:invmassXiStar} for 
$\Sigma^{*+}$$\rightarrow$ $\Lambda\pi^+$ and $\Xi^{*0}$($\overline{\Xi}^{*0}$) 
$\rightarrow$ $\Xi^-\pi^+$($\Xi^+\pi^-$), respectively. 
\footnote{Similarly to what has been observed in the pp analysis \cite{cite:Xi_pp}, the distributions of $\Sigma^{*-}$ 
($\overline{\Sigma}^{*+}$), not shown in this paper, have an additional peak at $\sim$1.321~\Gmass, as narrow as $\sim$3~\mmass, due 
to the residual $\Xi(1321)^{-}$ ($\overline{\Xi}(1321)^{+}$), escaping the filter on the DCA of $\Lambda$ to PV mentioned above.}

Since the resonance decay products originate from a position which is indistinguishable from the PV,
a significant combinatorial background is present. These background distributions were determined by means of 
a mixed-event technique, by combining uncorrelated decay products from 5 and 20 different events in 
the $\Sigma^{*\pm}$ and  $\Xi^{*0}$ analyses, respectively. In order to minimise distortions due to different 
acceptances and to ensure a similar event structure, only tracks from events with similar vertex 
positions $z$ ($|\Delta z| <$ 1 cm) and track multiplicities $n$ ($|\Delta n|<$ 10) were taken.

For $\Sigma^{*\pm}$, the mixed-event background distributions were normalised to a 
$\pt$-dependent invariant mass region where the mixed-event background and the invariant mass distribution have similar 
slopes, as shown in Fig.~\ref{fig:invmassSigmaStarPlus}~(Left). These $\pt$-dependent invariant mass regions range from  
1.5~$<$~$M_{\Lambda\pi}$~$<$~2.0~\Gmass, for the lowest $\pt$~bin, 
to 1.95$<$~$M_{\Lambda\pi}$~$<$~2.0~\Gmass, for the highest $\pt$~bin.
More details on the normalisation procedure are provided in Ref.~\cite{cite:Xi_pp}. The contribution of the 
normalisation to the systematic uncertainty was estimated by selecting different normalisation regions and 
accounts for less than 1\%. 

For $\Xi^{*0}$, the mixed-event background distributions were normalised to two fixed regions, 
\linebreak 1.49~$<$~$M_{\Xi\pi}$~$<$~1.51~\Gmass~and 1.56$<$~$M_{\Xi\pi}$~$<$~1.58~\Gmass,
~around the $\Xi^{*0}$ mass peak (Fig.~\ref{fig:invmassXiStar} (Left)). These regions were used 
for all $\pt$ intervals and multiplicity classes, because the background shape is reasonably well 
reproduced in these regions and the invariant-mass resolution of the reconstructed peaks appears stable, 
independently of $\pt$. The uncertainty on the normalisation was estimated by varying the normalisation regions and is 
included in the quoted systematic uncertainty for the signal extraction (Table~\ref{tab:sys}).

For $\Sigma^{*\pm}$, a combined fit of a second-order polynomial for the residual background 
description and a Breit-Wigner function with a width fixed to the PDG values~\cite{cite:PDG} for the signal 
were used in the invariant-mass range of $1.28<M_{\Lambda\pi}<1.55$~\Gmass. 
The detector resolution ($\sim$1~\mmass) is much lower than the $\Sigma^{*\pm}$ 
width and was therefore neglected. In the right panel of 
Fig.~\ref{fig:invmassSigmaStarPlus}, the solid and dashed lines show the result of the combined 
fit and the residual background, respectively.  Alternative fit ranges were taken into account in the estimation 
of the systematic uncertainty. A linear and a cubic parametrization for the residual background were used to 
study the systematic uncertainty related to the signal extraction. 

For $\Xi^{*0}$, a combined fit of a first-order polynomial for the residual background and a Voigtian function 
(a convolution of a Breit-Wigner and a Gaussian function accounting for the detector resolution) for the signal was 
used, as described in Ref.~\cite{ cite:Xi_pp}. 

The raw yields $N^{\rm RAW}$ were obtained by integrating the signal function from the combined fit.
For $\Sigma^{*\pm}$, the integration of the Breit-Wigner function was carried out in the invariant mass range 
between \linebreak1.28~\Gmass~and 1.56~\Gmass. For $\Xi^{*0}$, 
the integration of the Voigtian function was done in the mass region between 1.48~\Gmass~and 1.59~\Gmass. In both cases,
corrections for the tails outside the integration region were applied. The statistical uncertainties on the raw yields range 
between 5--15\% for $\Sigma^{*\pm}$ and 2--6\% for $\Xi^{*0}$, respectively. 

\subsection{Corrections and normalisation}
\begin{figure}[bt!bh]
  \begin{minipage}[c]{\linewidth}
    \subfigure{
      \resizebox{0.15\textwidth}{!}{
      }  
    \resizebox{.7\textwidth}{!}{
        \includegraphics[width=7.5cm]{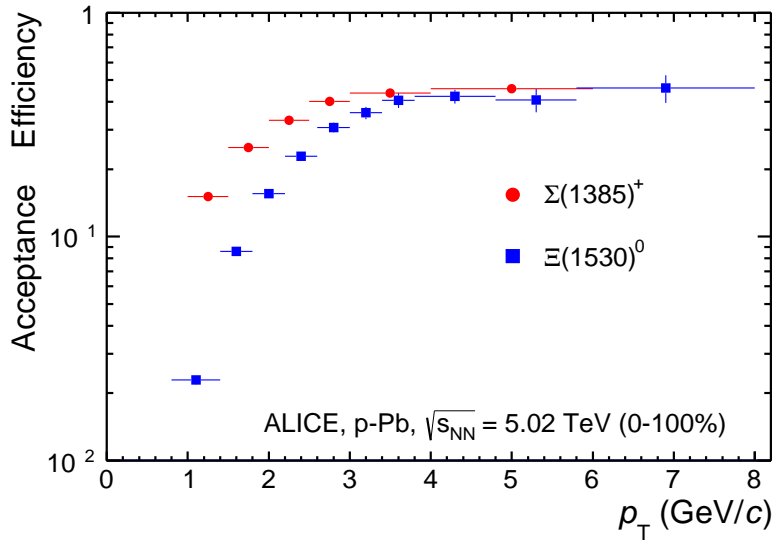}      
     }
    } 
  \caption{The geometrical acceptance and the reconstruction efficiency
  (A $\times$ $\epsilon$) for $\Sigma^{*+}$ and $\Xi^{*0}$ in $-0.5 < y_{\mathrm{CMS}^{\mathrm{MC}}} <$ 0 for 
  minimum-bias events, obtained with DPMJET~3.05~\cite{cite:DPMJET} and GEANT~3.1
  ~\cite{cite:GEANT}. Only statistical uncertainties are shown.}
  \label{fig:efficiency}
  \end{minipage}
\end{figure}

The raw yields were corrected for the geometrical acceptance and the reconstruction efficiency \break
(A $\times$ $\epsilon$) of the detector (Fig.~\ref{fig:efficiency}) and by branching ratios 
(total B.R. in Table~\ref{tab:PDG}). By using the DPMJET 3.05 event generator~\cite{cite:DPMJET} 
and the GEANT~3.21 package~\cite{cite:GEANT}, a sample of about 100 million p--Pb events was 
simulated and reconstructed in order to compute the corrections. The distributions of $A\times\epsilon$ 
were obtained from the ratio between the number of reconstructed hyperons ($\Sigma^{*\pm}$ or 
$\Xi^{*0}$) and the number of generated hyperons in the same \pt~and rapidity interval. 
Inefficiencies in the vertex reconstruction have a negligible effect for all multiplicity classes except 60-100\%, 
where a correction factor of 1.03 has to be applied to the raw yields.

The product $A\times\epsilon$ for MB events is shown in Fig.~\ref{fig:efficiency} for $\Sigma^{*+}$ and 
$\Xi^{*0}$. Since the correction factors for different multiplicity classes are in agreement with those from 
MB events within statistical uncertainty, the latter were used for all multiplicity classes. For $\Sigma^{*+}$ 
and  $\Sigma^{*-}$, the correction factors were the same. In the case of $\overline{\Sigma}^{*+}$ and 
$\overline{\Sigma}^{*-}$, correction factors were around 10\% higher at low $\pt$, as expected due to the different 
interaction cross sections of proton and antiprotons in the detector's material ~\cite{cite:antiproton}. 

Finally, the yields were normalised to the number of events analysed in each multiplicity class, as defined 
in Table \ref{tab:v0estimator}. The MB spectra were instead normalised to the number of 
NSD events after applying the correction factors for trigger efficiency and event 
selection, primary vertex reconstruction and selection, resulting in a total scaling factor of 
0.964~\cite{cite:KphipPb}.

\subsection{Systematic uncertainties}
Systematic effects due to the global tracking efficiency, track and topological selection cuts, PID, 
mass window selection ($\Xi^\pm$), vertex selection, signal extraction and uncertainties on the 
knowledge of the material budget and branching ratio were studied for each \pt~interval and multiplicity 
class by comparing different choices of selection criteria. The results are summarized in Table~\ref{tab:sys}.
\begin{table}[h!]
\centering
\begin{tabular}{lcc}
\hline\noalign{\smallskip}
Source of uncertainty & $\Sigma^{*\pm}$ & $\Xi^{*0}$ \\

\hline\noalign{\smallskip}
\pt-dependent & & \\
\hline\noalign{\smallskip}
Tracking efficiency & 3\% & 3\% \\
Tracks selection & 1-2\% &  1-2\% \\
Topological selection & 1-4\% & 1-2\% \\
PID  & 1-3\% &  3-7\% \\ 
Signal extraction & 2-5\% & 1-5\% \\
Mass window ($\Xi^\pm$)& - & 4\% \\
Vertex selection & 1-2\% & 3\% \\
\hline\noalign{\smallskip}
\pt-independent & & \\
\hline\noalign{\smallskip}
Material budget  & 4\% &  4\% \\
Branching ratio  & 1.1\% & 0.3\% \\
\hline\noalign{\smallskip}
Total & 7-9\% & 8-12 \% \\
\hline\noalign{\smallskip}
\end{tabular}
\caption{Summary of the systematic uncertainties on the differential yield, \dndydpt. 
Minimum and maximum values in all \pt~intervals and multiplicity classes are shown for 
each source.}
\label{tab:sys}    
\end{table}

Each source of systematic effects was first requested to pass a consistency check, testing whether 
a change in selection criteria prevents statistically significant differences in the reconstructed 
yields~\cite{cite:barlow}. If the source failed the consistency check, the deviation between the default 
yield and the alternative one obtained by varying the selection was taken as systematic uncertainty. 
Sources which did not provide statistically significant differences are not listed in Table~\ref{tab:sys} 
(e.g. $\Lambda$ invariant mass window). The uncertainty for the
$\Sigma^{*\pm}$ yield is taken as the average of the uncertainties for $\Sigma^{*+}$,
$\overline{\Sigma}^{*-}$, $\Sigma^{*-}$, and $\overline{\Sigma}^{*+}$.

For $\Sigma^{*\pm}$, the main contribution to the total systematic uncertainty originates from the signal extraction, 
while for $\Xi^{*0}$ the main contribution is from the PID. The signal extraction includes variations of the 
background normalisation region, choice of the integration interval of the raw yield determination and, in the case of 
$\Sigma^{*\pm}$, order of the polynomial for describing the residual background. Also, an alternative method, which integrates 
the signal distribution by summing the bin contents, provides negligible differences.

Table~\ref{tab:sys} reports the minimum and maximum of the systematic uncertainty from each source. 
The systematic uncertainty in each \pt~interval is obtained as the quadratic sum of all contributions, 
except the \pt-independent uncertainties, which affect only the normalisation (see Section~\ref{subsec:spectra}). 
The uncertainties which are dependent on multiplicity and uncorrelated across different multiplicity bins were 
treated separately. Topological selections, signal extraction and PID give 
the dominant contributions to the uncertainties uncorrelated across multiplicity. These uncertainties 
were estimated to be within 3\% (5\%), which represents a fraction of 35\% (50\%) of the total 
systematic uncertainty for $\Sigma^{*\pm}$ ($\Xi^{*0}$).

\section{Results and discussion}
\label{sec:results}
\subsection{Transverse momentum spectra}
\label{subsec:spectra}

\begin{figure}[b!htp]
  \begin{minipage}[c]{\linewidth}
    \subfigure{
     \resizebox{.5\textwidth}{!}{
        \includegraphics[width=7.5cm]{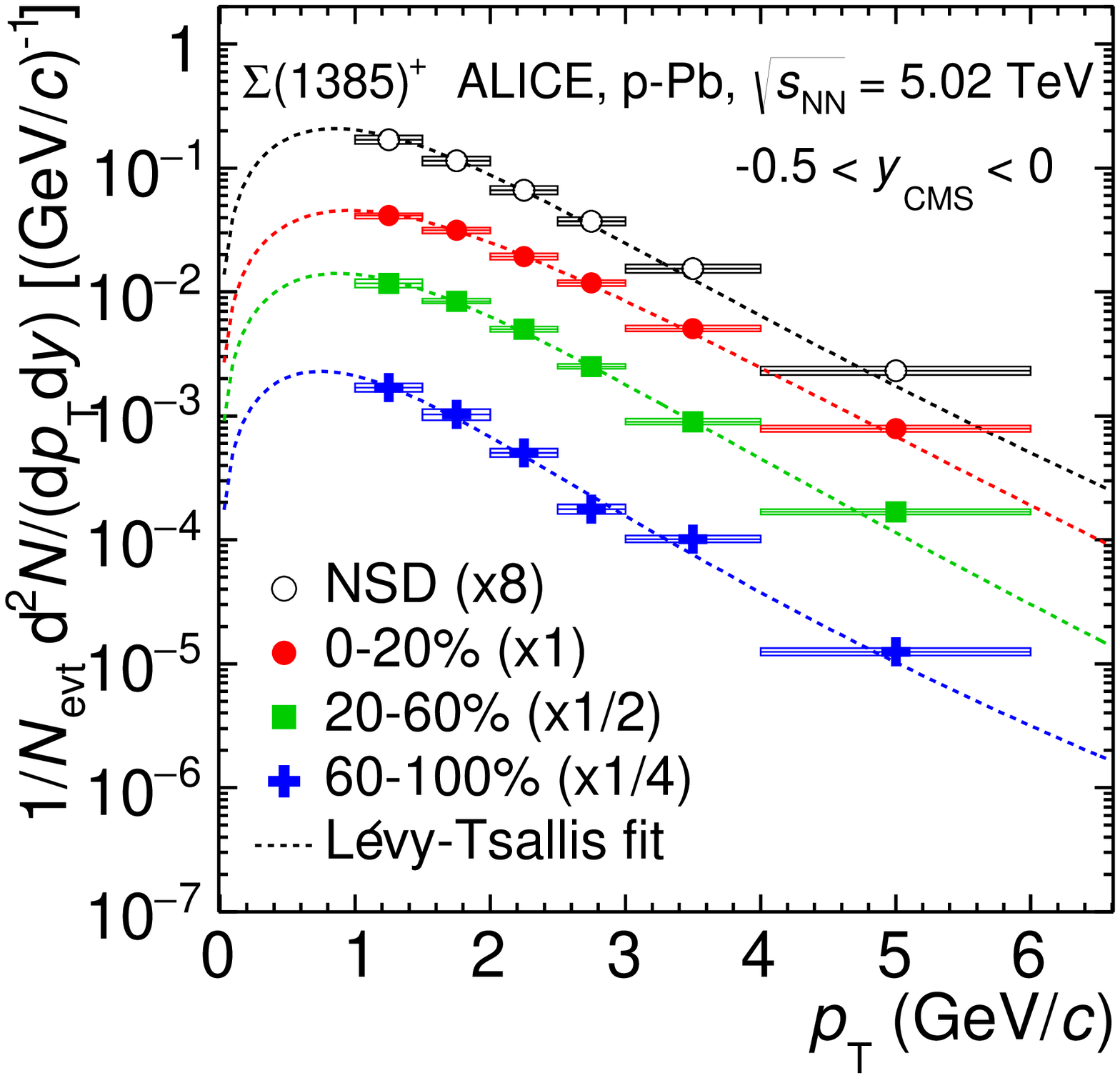}
     }
    } 
    \subfigure{
     \resizebox{.5\textwidth}{!}{
       \includegraphics[width=7.5cm]{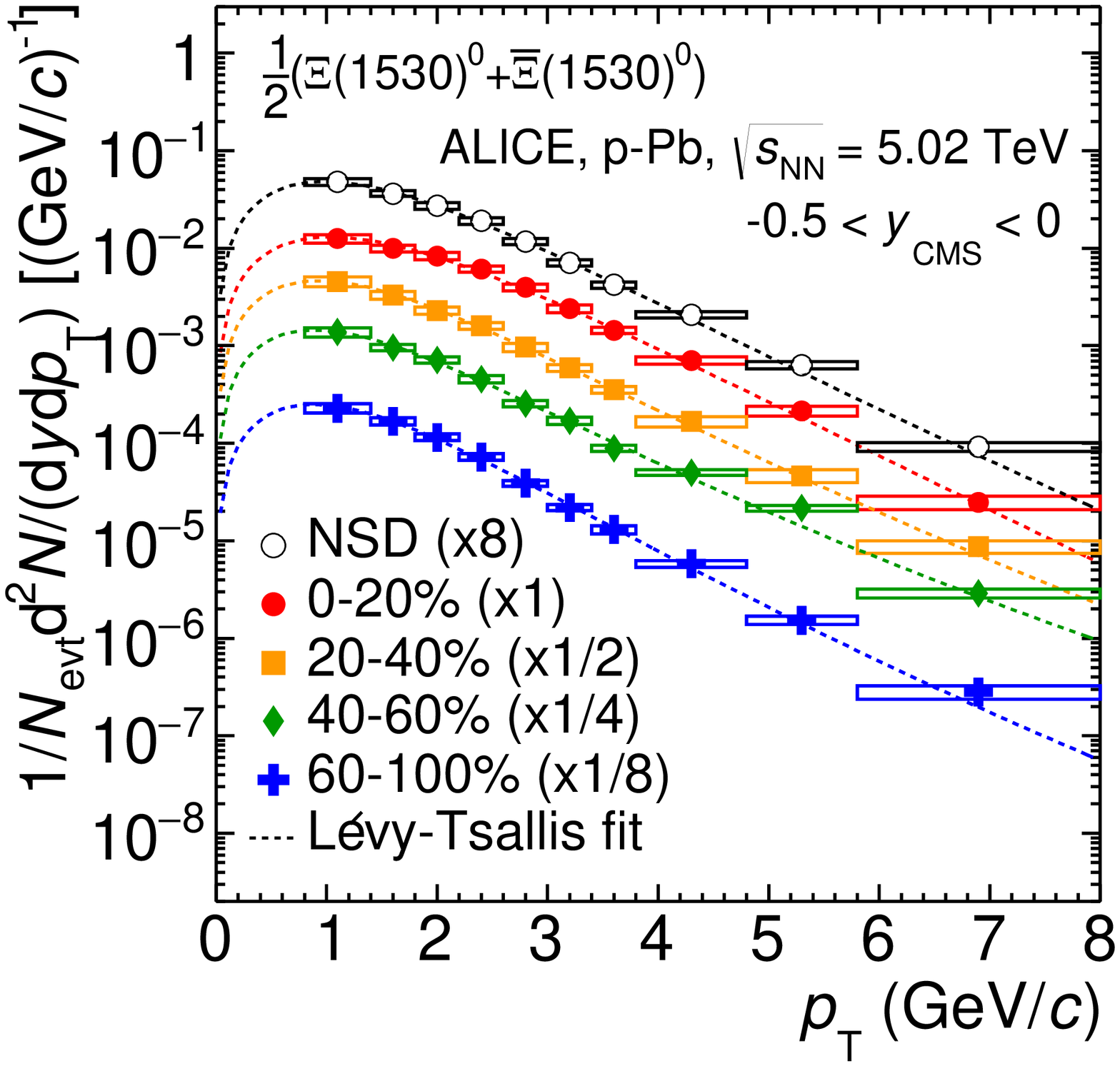}
     }
    }
  \caption{Transverse momentum spectra of $\Sigma^{*+}$ (left) and $\Xi^{*0}$ (right) in different multiplicity classes in 
  the rapidity range $-0.5<y_{\mathrm{CMS}}<0$. For~$\Xi^{*0}$, both particles and antiparticles 
  are analysed together. Statistical (bars) and systematic (boxes) uncertainties are included. The dashed curves 
  are L\'{e}vy-Tsallis fit to each individual distribution.}
 \label{fig:spectra}
  \end{minipage}
\end{figure}

The transverse momentum spectra of $\Sigma^{*+}$ and $\Xi^{*0}$ in the rapidity range 
$-0.5<y_{\mathrm{CMS}}<\nobreak0$ are shown in Fig.~\ref{fig:spectra} for different multiplicity classes and for 
NSD events. They cover the ranges 1~$<$~\pt~$<$~6~\gmom~for $\Sigma^{*+}$ 
and 0.8~$<$~\pt~$<$~8~\gmom~for $\Xi^{*0}$. The spectra obtained for 
$\overline{\Sigma}^{*-}$, $\Sigma^{*-}$ and $\overline{\Sigma}^{*+}$ are 
consistent with the spectrum of $\Sigma^{*+}$.

The spectra are fitted with a L\'{e}vy-Tsallis function~\cite{cite:Tsallis}, 
\begin{equation}
\frac{1}{N_{\mathrm{evt}}}\frac{\mathrm{d}^2N}{\mathrm{d}p_{\mathrm{T}}\mathrm{d}y} = p_{\mathrm{T}} \frac{\mathrm{d}N}{\mathrm{d}y} \frac{(n-1)(n-2)}{nC[nC+m_{0}(n-2)]}\left[1+\frac{\sqrt{p_{\mathrm{T}}^2+m_{0}^{2}}-m_{0}}{nC}\right]^{-n},\label{eqn:funclevy}
\end{equation}
where $N_\mathrm{evt}$ is the number of events, $m_{0}$ is the mass of the particle, and $n$, $C$ and the 
integrated yield \dndy~are free parameters for the fit.
This function was successfully used to describe most of the identified particle spectra in pp collisions~
\cite{cite:Xi_pp,cite:KphipPb, cite:lambda_pp}.

The values of \dndy~and $\mpt$ shown in Table~\ref{tab:summary} were calculated by using the experimental 
spectrum in the measured \pt-range and the L\'{e}vy-Tsallis fit function outside of the measured \pt-range. 
The contribution from the low-$\pt$ extrapolation to the total \dndy~is 
36-47\% (20-29\%) for $\Sigma^{*+}$ ($\Xi^{*0}$) moving from low to high multiplicity, while the one from the 
high-$\pt$ extrapolation is negligible. The systematic uncertainties on \dndy~and $\mpt$ presented in 
Table~\ref{tab:summary} were estimated by repeating the L\'{e}vy-Tsallis fit moving 
randomly (with a Gaussian distribution) the measured points within their $\pt$-dependent systematic uncertainties.
The $\pt$-independent uncertainties were further added in quadrature to the systematic 
uncertainties on \dndy.  Alternative functional forms, such as Boltzmann-Gibbs Blast-Wave
~\cite{cite:blastwave, cite:STAR-ratio_to_pion}, $m_{\rm{T}}$-exponential~\cite{cite:STAR-Kstar-2005, 
cite:STAR-ratio_to_pion}, Boltzmann and Bose-Einstein fit functions were used for both particles 
to evaluate the systematic uncertainties on the low-$\pt$~extrapolation.
The maximum difference between the results obtained with the various fit functions was taken as 
the uncertainty. These systematic uncertainties, which vary between 5\% and 10\%, were added in quadrature to the 
uncertainties for the L\'{e}vy-Tsallis fit. The values for 
$\Sigma^{*\pm}$ in Table~\ref{tab:summary}~were obtained by averaging those for 
$\Sigma^{*+}$, $\overline{\Sigma}^{*-}$, $\Sigma^{*-}$ and 
$\overline{\Sigma}^{*+}$ to reduce the statistical uncertainties. \\

\begin{table}[h!]
\centering
\begin{tabular}{c|c|c|c}
\hline\noalign{\smallskip}
Baryon & Multiplicity class & \dndy~{\scriptsize ($\times$10$^{-3}$)} &  $\mpt$ (\gmom) \\
\hline\noalign{\smallskip}
\multirow{4}{*}{$\Sigma^{*\pm}$} 
& NSD               &  49.0 $\pm$ 0.6  $\pm$ 6.5   &  1.367 $\pm$ 0.009 $\pm$ 0.061  \\
& 0-20\%           &  90.3 $\pm$ 1.4 $\pm$ 7.9   &  1.495 $\pm$ 0.012 $\pm$ 0.046  \\
& 20-60\%         &  52.2 $\pm$ 0.8 $\pm$ 6.0   &  1.342 $\pm$ 0.010 $\pm$ 0.055  \\
& 60-100\%       &  15.2 $\pm$ 0.4 $\pm$ 2.4   &  1.173 $\pm$ 0.015 $\pm$ 0.067  \\
\hline
\multirow{5}{*}{1/2($\Xi^{*0}+\overline{\Xi}^0$)} 
& NSD               & 12.5 $\pm$ 0.3 $\pm$ 1.1 & 1.540 $\pm$ 0.016 $\pm$ 0.071 \\
& 0-20\%            & 27.3 $\pm$ 0.6 $\pm$ 2.8 & 1.626 $\pm$ 0.016 $\pm$ 0.068 \\
& 20-40\%           & 17.7 $\pm$ 0.5 $\pm$ 2.4 & 1.482 $\pm$ 0.020 $\pm$ 0.100 \\
& 40-60\%           & 10.7 $\pm$ 0.3 $\pm$ 1.6 & 1.459 $\pm$ 0.025 $\pm$ 0.114 \\
& 60-100\%          &  3.6 $\pm$ 0.1 $\pm$ 0.5 & 1.377 $\pm$ 0.023 $\pm$ 0.089 \\
\noalign{\smallskip}\hline\noalign{\smallskip} 
\end{tabular}
\caption{Integrated yields (\dndy) and mean transverse 
momenta ($\mpt$). The values for $\Sigma^{*\pm}$
are obtained by averaging the values for $\Sigma^{*+}$, 
$\overline{\Sigma}^{*-}$, $\Sigma^{*-}$ and $\overline{\Sigma}^{*+}$.
Statistical (first one) and total systematic (second one) uncertainties including the extrapolation from the various fit functions are quoted.}
\label{tab:summary}     
\end{table}

%
%

\subsection{Mean transverse momenta} \label{subsec:mean_pt}
\begin{figure}[h!]
  \begin{minipage}[c]{\linewidth}
    \subfigure{
     \resizebox{0.25\textwidth}{!}{
	}
     \resizebox{0.5\textwidth}{!}{
        \includegraphics[width=5.5cm]{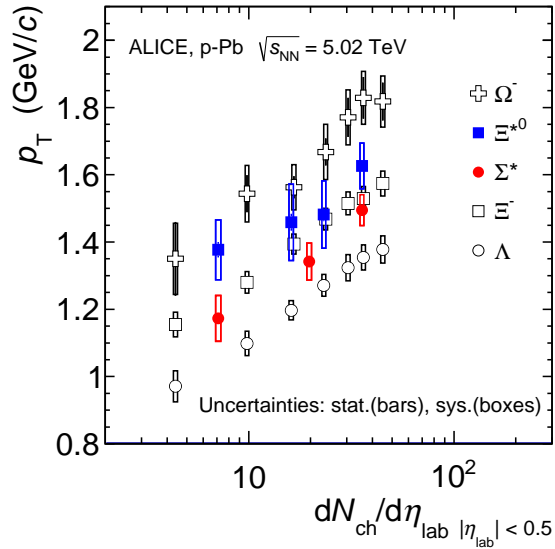}
     }
    } 
  \caption{Mean transverse momenta $\mpt$ of $\Lambda$, $\Xi^{-}$, $\Sigma^{*\pm}$, $\Xi^{*0}$ 
  and $\Omega^{-}$ in p--Pb collisions at $\snn$ = 5.02 TeV as a 
  function of mean charged-particle multiplicity density 
  $\langle$d$N_{\mathrm{ch}}$/d$\eta_{\mathrm{lab}}\rangle$, measured in the pseudorapidity range 
  $\mid\eta_{\mathrm{lab}}\mid <$~0.5. The results for $\Lambda$, $\Xi^{-}$ and $\Omega^{-}$ are taken 
  from~\cite{cite:lambda_pPb, cite:KphipPb, cite:Xi_pPb}. Statistical and systematic uncertainties are represented as bars 
  and boxes, respectively. The $\Omega^-$ and $\Xi^-$ points in the 3rd and 4th lowest multiplicity bins are slightly 
  displaced along the abscissa to avoid superposition with the $\Xi^{*0}$ points.}
  \label{fig:mean_pt_vs_mult}
  \end{minipage}
\end{figure}

Figure~\ref{fig:mean_pt_vs_mult} shows the mean transverse momentum $\mpt$ as a function of mean 
charged-particle multiplicity density $\langle$d$N_{\mathrm{ch}}$/d$\eta_{\mathrm{lab}}\rangle$ at midrapidity.
The results for $\Sigma^{*\pm}$ and $\Xi^{*0}$ are compared with those for other hyperons  
observed in p--Pb collisions at $\sqrt{s_{\rm NN}}$~=~5.02~TeV~\cite{cite:lambda_pPb, cite:Xi_pPb}.

Increasing trends from low to high multiplicities are observed for all hyperons. For both $\Sigma^{*\pm}$ 
and $\Xi^{*0}$, the mean transverse momenta increase by 20\% as the mean charged-particle multiplicity 
increases from 7.1 to 35.6. This result is similar to the one obtained for the other hyperons.
Furthermore, a similar increase has been observed also for K$^{\pm}$, K$_{\rm{S}}^{0}$, K$^{*}(892)^0$ 
and $\phi$~\cite{cite:KphipPb}, whereas protons are subject to a larger ($\sim$ 33\%) 
increase in the given multiplicity range, as discussed also in Ref.~\cite{cite:lambda_pPb}. 
\begin{figure}[h!]
  \begin{minipage}[c]{\linewidth}
    \subfigure{
     \resizebox{0.15\textwidth}{!}{
}
     \resizebox{0.6\textwidth}{!}{
        \includegraphics[width=10cm]{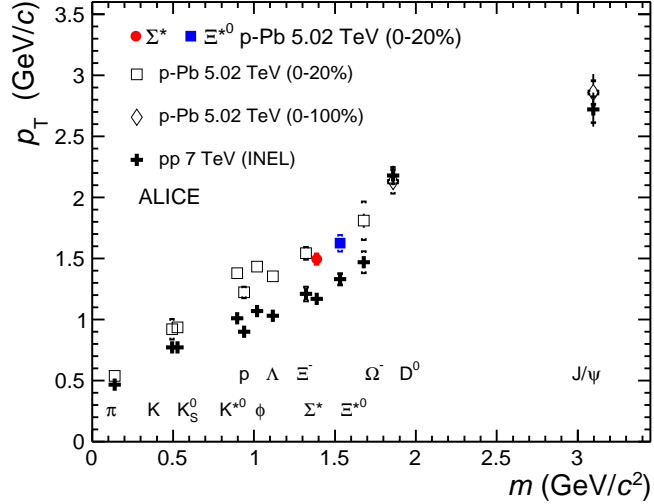}
     }
    } 
  \caption{Mass dependence of the mean transverse momenta of identified particles 
  for the $0-20$\% V0A multiplicity class and with 
  $-0.5<y_{\mathrm{CMS}}<0$ in p--Pb~collisions at 
  $\sqrt{s_{\mathrm{NN}}}$~=~5.02~TeV~\cite{cite:lambda_pPb, cite:Xi_pPb}, and 
  in minimum-bias pp collisions at $\sqrt{s}$~=~7~TeV~\cite{cite:Xi_pp} with 
  $|y_{\mathrm{CMS}}|<0.5$. Additionally, $D^0$ and $J$/$\psi$ results are plotted. 
  The $D^0$ and $J$/$\psi$ were measured in different rapidity ranges: $|y_{\mathrm{CMS}}|<0.5$~\cite{cite:D0} 
  ($|y_{\mathrm{CMS}}|<0.9$~\cite{cite:Jpsi_pp}) for $D^0$ ($J$/$\psi$) in pp and 
  $-0.96 < y_{\mathrm{CMS}}< 0.04$~\cite{cite:D0} ($-1.37<y_{\mathrm{CMS}}<0.43$~\cite{cite:Jpsi_pPb}) 
  for $D^0$ ($J$/$\psi$) in p--Pb. Note also that the results for $D^0$ and $J$/$\psi$ in p-Pb collisions 
  are for the 0-100\% multiplicity class.}
  \label{fig:mean_pt_vs_mass}
  \end{minipage}
\end{figure}

In all multiplicity classes, the $\mpt$ follows an approximate mass ordering: 
$\mpt_{\Lambda}<\mpt_{\Xi^-}$~$\simeq$~$\mpt_{\Sigma^{*\pm}}<\mpt_{\Xi^{*0}}<\mpt_{\Omega^{-}}$.
The $\mpt$ of $\Sigma^{*\pm}$ looks systematically lower than the $\mpt$ of $\Xi^{-}$, despite the larger mass of 
$\Sigma^{*\pm}$. The uncertainties, however, are too large to draw any conclusion on 
possible hints of violation of the mass hierarchy. This hierarchy of mass-ordering, also including $D^0$ and 
$J$/$\psi$ in the comparison, is displayed in Fig.~\ref{fig:mean_pt_vs_mass}. Note, however, that the $D^0$ and $J$/$\psi$ were 
measured in different rapidity ranges: $|y_{\mathrm{CMS}}|<0.5$~\cite{cite:D0} 
($|y_{\mathrm{CMS}}|<0.9$~\cite{cite:Jpsi_pp}) for $D^0$ ($J$/$\psi$) in pp and 
$-0.96 < y_{\mathrm{CMS}}< 0.04$~\cite{cite:D0} ($-1.37<y_{\mathrm{CMS}}<0.43$~\cite{cite:Jpsi_pPb}) 
for $D^0$ ($J$/$\psi$) in p--Pb, and the results for $D^0$ and $J$/$\psi$ in p-Pb collisions 
are for the 0-100\% multiplicity class. This mass dependence is observed in both p--Pb and pp collisions. 
It was observed also by the STAR collaboration~\cite{cite:STAR-hadronic_resonances-dAu} in MB pp, MB 
d--Au and central Au--Au collisions. 

Furthermore, for the light-flavour hadrons, the mean transverse momenta in p--Pb collisions are 
observed to be consistently higher than those in pp collisions at 7 TeV. The situation for the charm 
hadrons is different, where $\mpt$ appears compatible between both colliding systems. The discrepancy 
is likely due to different production mechanisms for heavy and light flavours and to a harder fragmentation 
of charm quarks. Specifically, the fact that $\mpt$ remains similar in pp and in p--Pb is consistent with 
(i) the fact that p--Pb collisions can be considered as a superposition of independent nucleon-nucleon collisions 
for what concerns $D$-meson production, as described in \cite{cite:D0}, and/or 
(ii) with the effects of shadowing in p--Pb which reduces the production at low $\pt$ and thus increasing 
the overall $\mpt$ for $J$/$\psi$~\cite{cite:Jpsi_pPb}; the small $\pt$ hardening expected in pp when 
going from 5.02 to 7 TeV is apparently not enough to counter-balance the situation.

Because of small decrease of the $\mpt$ for proton and $\Lambda$ relative to those for K$^{*0}$ and $\phi$, 
two different trends for mesons and baryons have been suggested~\cite{cite:mass_scaling}. 
Even including $D^0$ and $J$/$\psi$, as shown in Fig.~\ref{fig:mean_pt_vs_mass}, a different trend for 
mesons and baryons cannot be convincingly established.

\subsection{Integrated particle ratios}  \label{subsec:yield_ratio}
\begin{figure}[t!]
  \begin{minipage}[c]{\linewidth}
    \subfigure{
     \resizebox{.5\textwidth}{!}{
        \includegraphics[width=7.5cm]{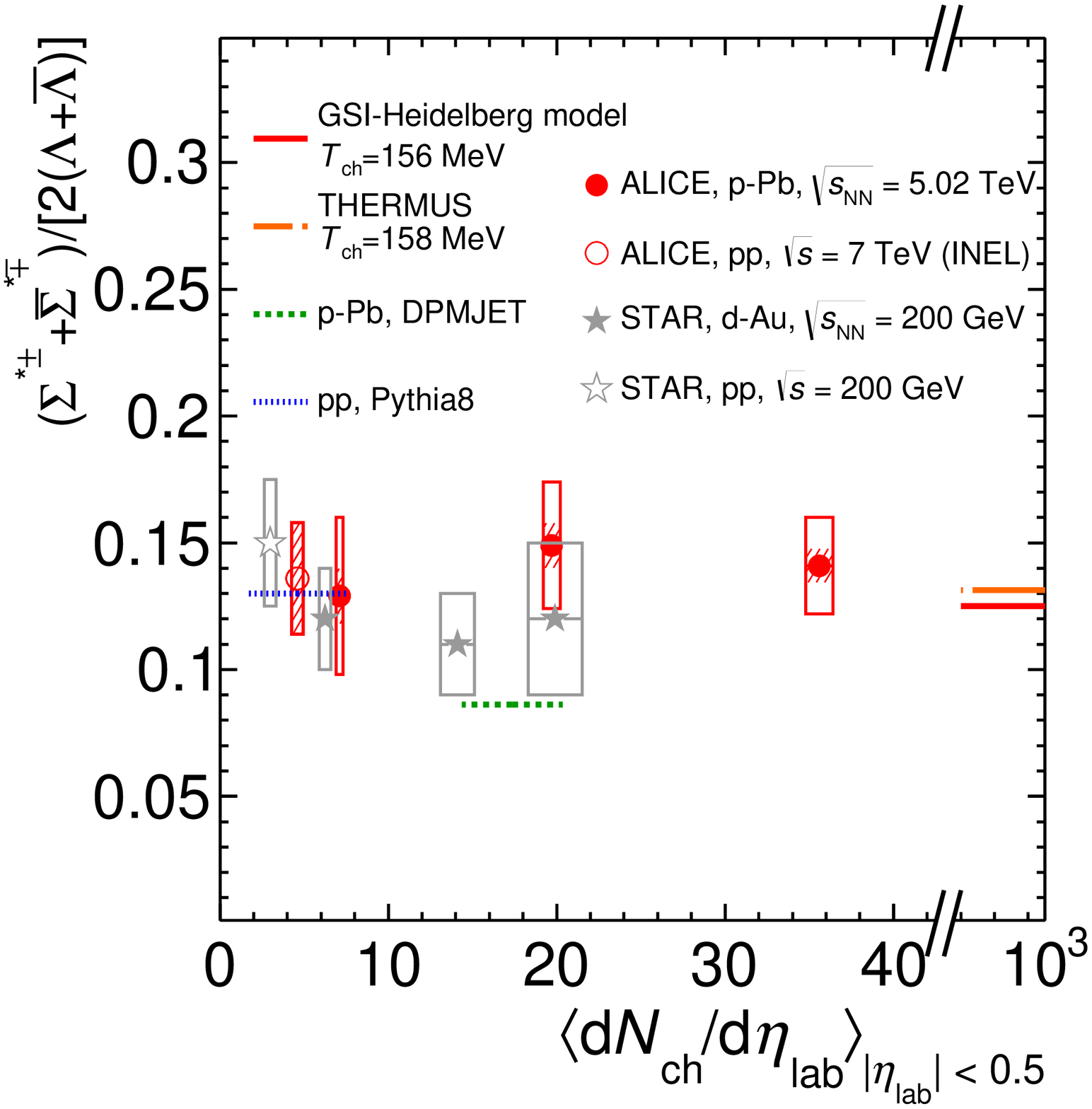}
     }
    } 
    \subfigure{
     \resizebox{.5\textwidth}{!}{
      \includegraphics[width=7.5cm]{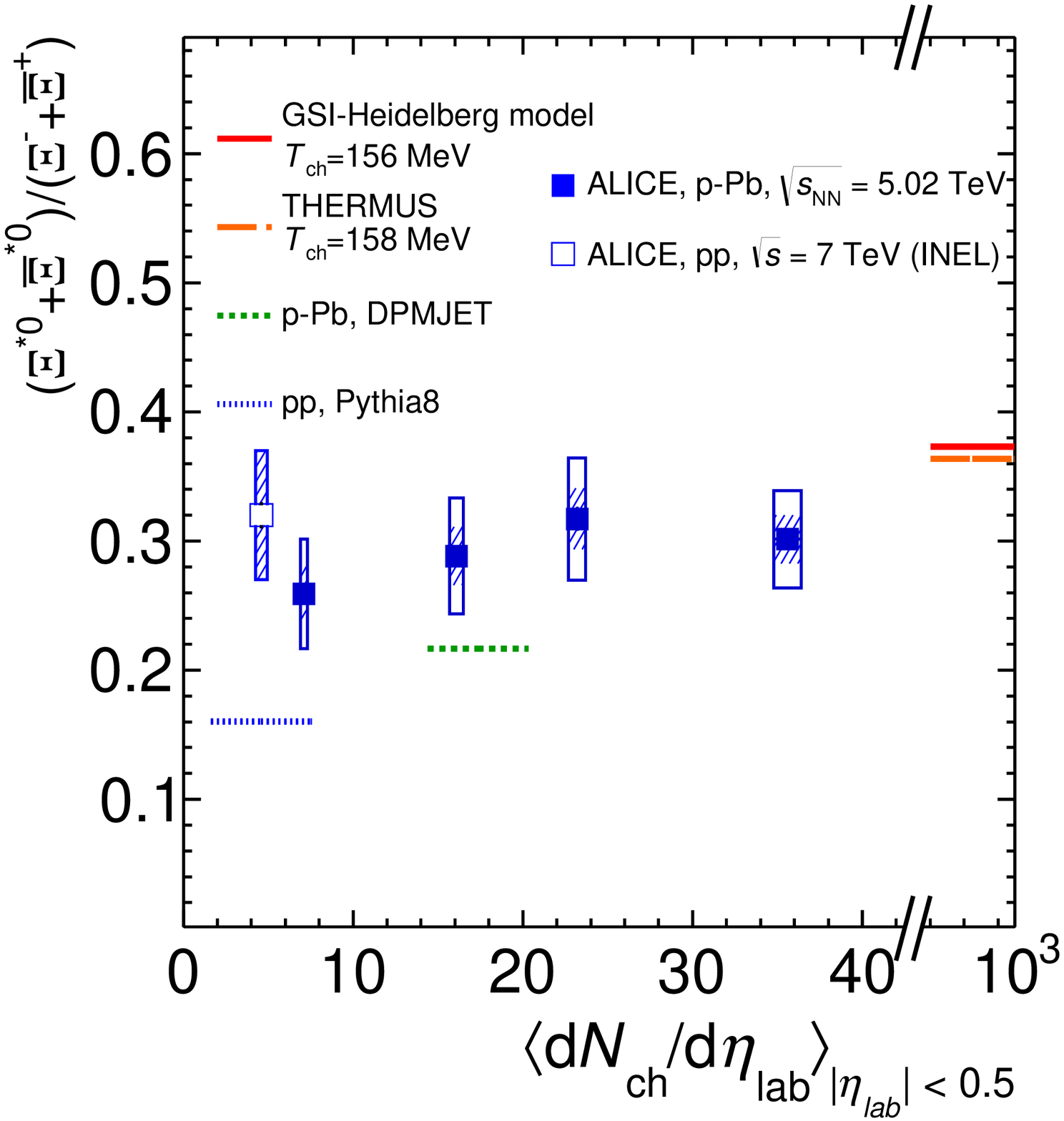}
     }
    } 
  \caption{(Left) Ratio of $\Sigma^{*\pm}$ to $\Lambda$ and (Right) ratio of $\Xi^{*0}$ 
  to $\Xi^-$ measured in \pp~\cite{cite:STAR-ratio_to_pion, cite:pp7_multistrange, cite:Xi_pp, cite:STAR-hadronic_resonances-dAu}, 
  d--Au~\cite{cite:STAR-ratio_to_pion, cite:STAR-hadronic_resonances-dAu} and 
  \pPb~\cite{cite:lambda_pPb, cite:Xi_pPb} collisions, 
  as a function of $\langle$d$N_{\mathrm{ch}}$/d$\eta_{\mathrm{lab}}\rangle$ measured at midrapidity.  
  Statistical uncertainties (bars) are shown as well as total systematic uncertainties (hollow boxes) and systematic 
  uncertainties uncorrelated across multiplicity (shaded boxes).
  A few model predictions are also shown as lines at their appropriate abscissa.}
  \label{fig:ratio_to_strange}
  \end{minipage}
\end{figure}

The integrated yield ratios of excited to ground-state hyperons 
~\cite{cite:pp7_multistrange, cite:lambda_pPb, cite:Xi_pp, cite:Xi_pPb, cite:STAR-ratio_to_pion, 
cite:STAR-hadronic_resonances-dAu} 
with the same strangeness content, for different collision systems and energies, are shown in 
Fig.~\ref{fig:ratio_to_strange} as a function of $\langle$d$N_{\mathrm{ch}}$/d$\eta_{\mathrm{lab}}\rangle$. 
In both cases, the variation of the integrated yield ratio with mean multiplicity is within experimental uncertainties.
In fact, the similar flat behaviour of $\Sigma^{*\pm}/\Lambda$ and $\Xi^{*0}/\Xi^-$ is remarkable, 
when considering their different lifetimes and other properties such as spin and mass.

\begin{figure}[h!]
  \begin{minipage}[c]{\linewidth}
    \subfigure{
     \resizebox{.5\textwidth}{!}{
        \includegraphics[width=7.5cm]{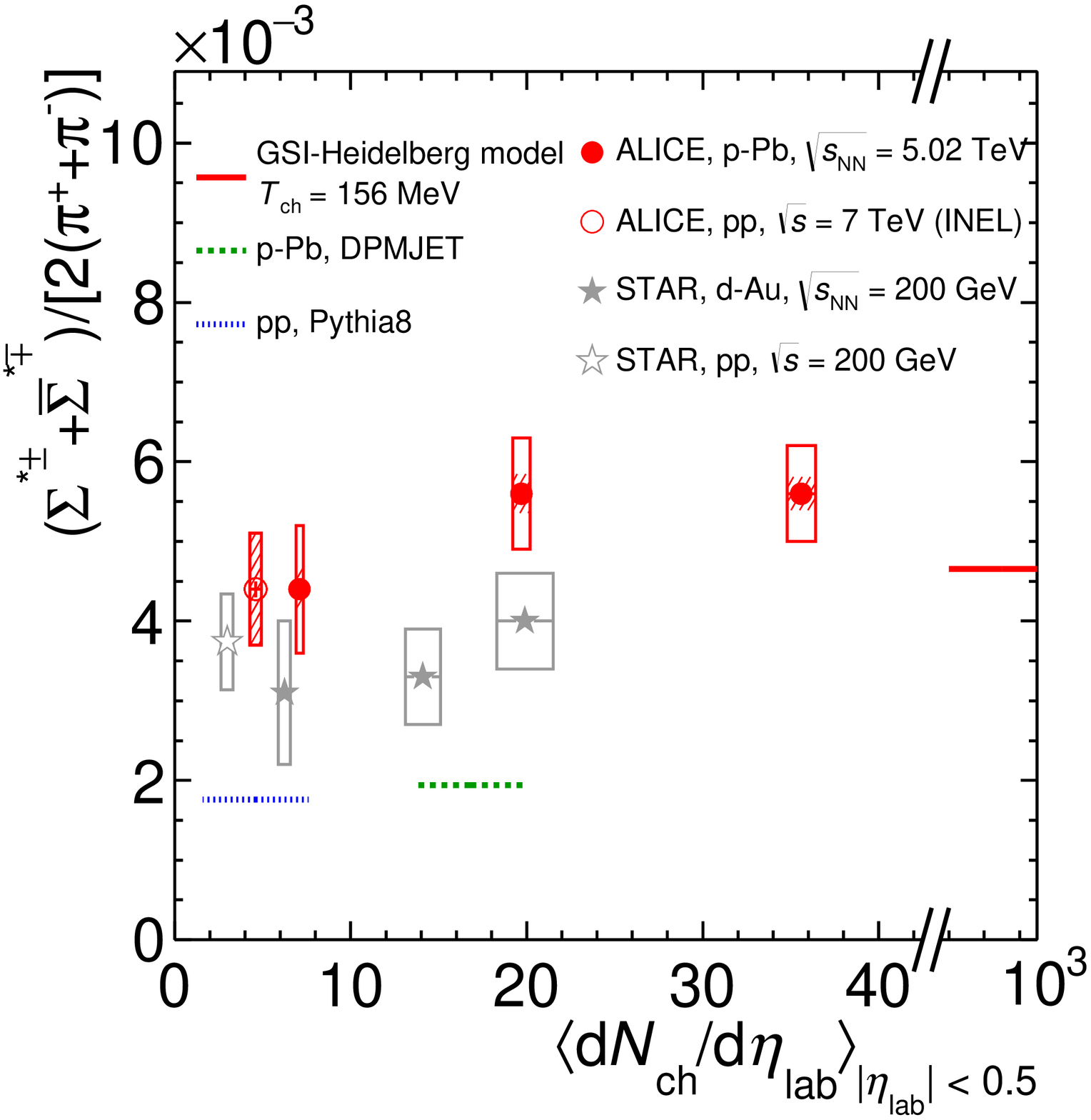}
     }
    } 
    \subfigure{
     \resizebox{.5\textwidth}{!}{
      \includegraphics[width=7.5cm]{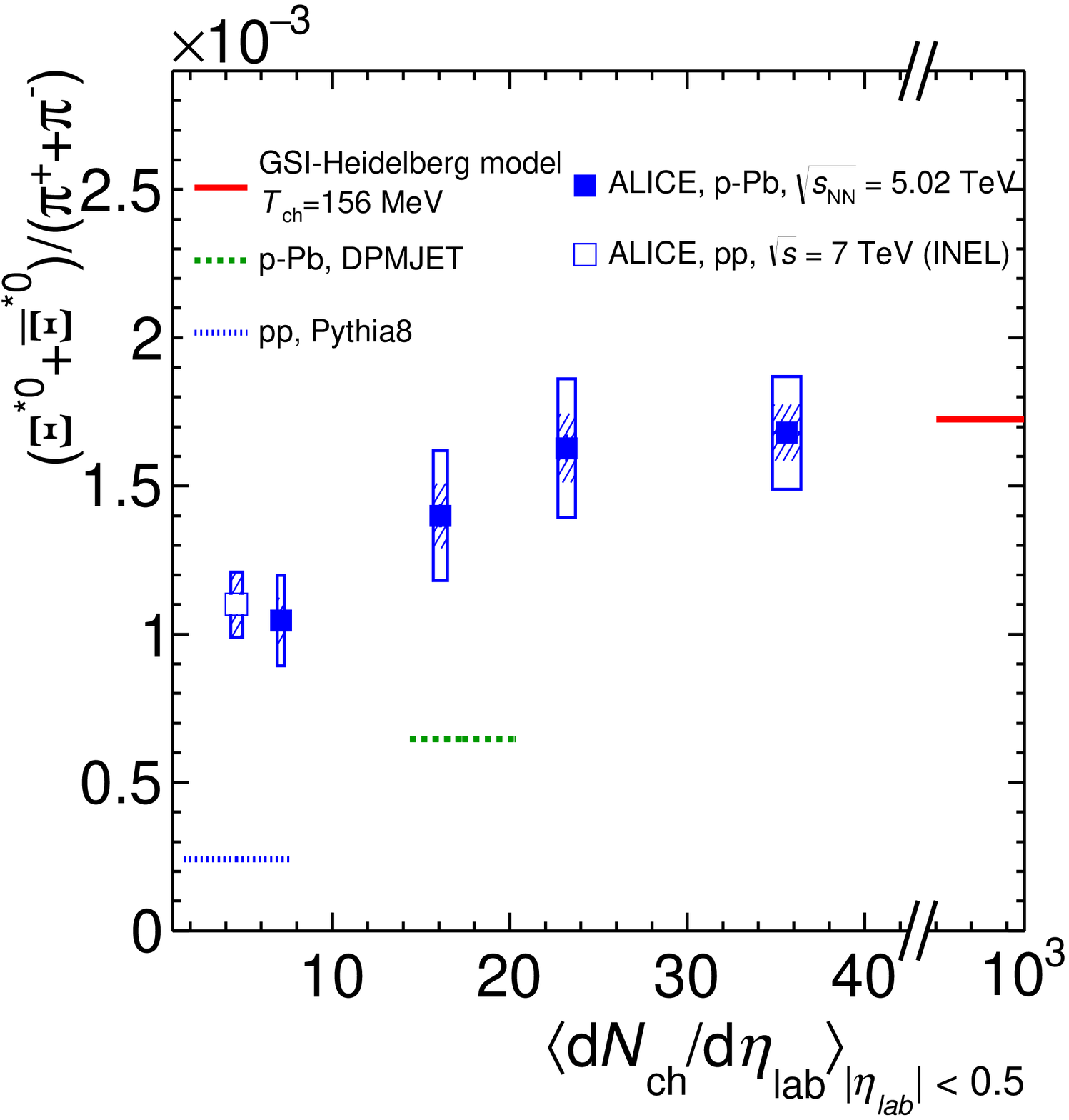}
     }
    } 
  \caption{(Left) Ratio of $\Sigma^{*\pm}$ to $\pi^{\pm}$ and (Right) ratio of $\Xi^{*0}$ to $\pi^{\pm}$, measured 
  in \pp~\cite{cite:Xi_pp, cite:pp7_piKp, cite:STAR-ratio_to_pion, cite:STAR_Strange_Baryon}, 
  d--Au~\cite{cite:STAR-ratio_to_pion, cite:STAR-hadronic_resonances-dAu} 
  and \pPb~\cite{cite:lambda_pPb} collisions, 
  as a function of the average charged particle 
  density ($\langle$d$N_{\mathrm{ch}}$/d$\eta_{\mathrm{lab}}\rangle$) measured at midrapidity. 
  Statistical uncertainties (bars) are shown as well as total systematic uncertainties (hollow boxes) and systematic 
  uncertainties uncorrelated across multiplicity (shaded boxes). A few model predictions are also shown as lines at their appropriate abscissa.}
  \label{fig:ratio_to_pion}
  \end{minipage}
\end{figure}

The results are compared with model predictions, PYTHIA8 for \pp~at 7 TeV~\cite{cite:pythia8} and DPMJET 
for \pPb~at 5.02 TeV~\cite{cite:DPMJET} collisions. The $\Sigma^{*\pm}$/$\Lambda$ ratios are consistent with 
the values predicted by PYTHIA8 in~\pp~collisions, whereas the DPMJET prediction 
for~\pPb~collisions is lower than the experimental data. The measured $\Xi^{*0}$/$\Xi^-$ 
ratios appear higher than the corresponding predictions for both systems. 
Note that the PYTHIA8~\cite{cite:pythia8} and DPMJET~\cite{cite:DPMJET} 
values in Figs.~\ref{fig:ratio_to_strange} and~\ref{fig:ratio_to_pion} were obtained respectively for INEL pp
and NSD p--Pb events, which have corresponding mean charged-particle multiplicities of  
$\langle$d$N_{\mathrm{ch}}$/d$\eta_{\mathrm{lab}}\rangle_\mathrm{INEL}$ = 4.60 $^{+0.34}_{-0.17}$ 
~\cite{cite:pp_mul} and $\langle$d$N_{\mathrm{ch}}$/d$\eta_{\mathrm{lab}}\rangle_\mathrm{NSD}$
 = 17.4 $\pm$ 0.7 ~\cite{cite:rapidity}. These predictions are indicated as dotted and dashed lines with arbitrary lengths 
 in the pertinent multiplicity regions in Figs.~\ref{fig:ratio_to_strange} 
 and~\ref{fig:ratio_to_pion}. Fig.~\ref{fig:ratio_to_pion} will be discussed later.

The results are also compared to thermal model predictions~\cite{cite:PbPb-GSI,cite:pp-THERMUS}. For small 
systems a canonical treatment is a priori required to take into account exact strangeness 
conservation~\cite{cite:pp-THERMUS}. This approach leads to a dependence on system size as can be 
seen in p-Pb collisions studying multi-strange hadrons~\cite{cite:Xi_pPb}. For the chosen ratios, however, 
the canonical corrections are identical for numerator and denominator (same strangeness quantum number). 
Therefore, the grand canonical values are used in Fig.~\ref{fig:ratio_to_strange} for two 
models~\cite{cite:PbPb-GSI,cite:pp-THERMUS}, which are marked at the asymptotic limit, corresponding 
to the mean charged-particle multiplicity in Pb--Pb~\cite{cite:PbPb-MultCentrDep}.

The constant behaviour of the yield ratios of excited to 
ground-state hyperons with same strangeness content indicates that neither regeneration nor re-scattering dominates 
with increasing collision system size, even for $\Sigma^{*\pm}$, which has a shorter lifetime than $\Xi^{*0}$ by a factor 
of 4. It is especially interesting to consider the constant behaviour of $\Sigma^{*\pm}$
/$\Lambda$ ratio in contrast to the apparent decrease observed for K$^{*0}$/K$^-$ ratio in the same 
$\langle$d$N_{\mathrm{ch}}$/d$\eta_{\mathrm{lab}}\rangle$ range~\cite{cite:KphipPb}, in spite of the similarly
short lifetimes of $\Sigma^{*\pm}$ and K$^{*0}$. In Pb--Pb collisions, both behaviours are predicted by the EPOS3 model 
\cite{cite:EPOS3, cite:EPOS3_2}, which employs the UrQMD model~\cite{cite:UrQMD} for the description of the hadronic phase.
In addition, the $\Sigma^{*\pm}$/$\Lambda$ ratios at 
LHC energies turn out to be comparable with the results obtained at lower energies by the STAR 
collaboration~\cite{cite:STAR-ratio_to_pion, cite:STAR-hadronic_resonances-dAu}. 

The integrated yield ratios of excited hyperons to pions are shown in Fig.~\ref{fig:ratio_to_pion} to study the evolution 
of relative strangeness production yields with increasing collision system size. Considering the relatively small 
systematic uncertainties uncorrelated across multiplicity (shaded boxes), one observes increasing patterns by 
40-60\% relative to results in \pp~collisions at the same $\sqrt{s_{\mathrm{NN}}}$, depending on the 
strangeness contents. These results are consistent with previous observations of ground-state hyperons to pion 
ratios measured at ALICE~\cite{cite:Xi_pPb}. The constant behavior of the $\Sigma^{*\pm}$/$\Lambda$ and 
$\Xi^{*0}/\Xi^-$ ratios indicates that the strangeness enhancement observed in p-Pb collisions depends predominantly 
on the strangeness content, rather than on the hyperon mass. Results from low-energy 
collisions \cite{cite:STAR-ratio_to_pion, cite:STAR-hadronic_resonances-dAu, cite:STAR_Strange_Baryon} show a similar pattern
in spite of the narrower range accessible for mean charged-particle multiplicity. 
In both cases, QCD-inspired predictions like PYTHIA for pp~\cite{cite:pythia8} and DPMJET for p--Pb~\cite{cite:DPMJET} clearly 
underestimate the observed yield ratios, while the statistical one seems to be comparable with results from 
high multiplicity events.

\section{Conclusions}
\label{sec:conclusion}

Transverse momentum spectra of~$\Sigma^{*\pm}$ and $\Xi^{*0}$ produced in \pPb~collisions at 
$\sqrt{s_{\rm NN}}$~=~5.02~TeV have been measured, and the yields and mean \pt~values have been extracted 
with the help of \Tsallis~fits. The mean $\pt$ of these hyperon resonances 
exhibit a similarly increasing pattern as other hyperons ($\Lambda$, $\Xi^-$, $\Omega^-$), depending on 
mean multiplicity and following the approximate mass ordering observed for other particles despite of
relatively large uncertainties. The integrated yield ratios of excited 
to ground-state hyperons, with the same strangeness content, show a flat behaviour over the whole mean multiplicity range. 
The $\Sigma^{*\pm}$/$\Lambda$ ratio does not show a variation with collision energy, nor with increasing system size.
The $\Xi^{*0}$/$\Xi^-$ ratios are higher than predicted by event generators. Both ratios agree with thermal 
model values. The yield ratios relative to pions show a gradual increase with 
$\langle$d$N_{\mathrm{ch}}$/d$\eta_{\mathrm{lab}}\rangle$. This rise is consistent with the results of 
ground-state hyperons produced in the same collision system, i.e. they show a gradual evolution with the system size 
depending only on the strangeness content. 

The current measurement represents a relevant baseline for further investigation in Pb--Pb collisions. It will be 
especially valuable to compare the  $\Sigma^{*\pm}$/$\Lambda$ ratio with K$^{*0}$/K$^-$, since $\Sigma^{*\pm}$
and K$^{*0}$ have similar lifetimes. A complete set of such measurements for many resonances ($\rho$, K$^{*0}$, $\phi$, 
$\Sigma^{*\pm}$, $\Lambda^*$, $\Xi^{*0}$) with different lifetimes will allow the properties of the hadronic phase to 
be studied in more detail.

%
\newenvironment{acknowledgement}{\relax}{\relax}
\begin{acknowledgement}
\section*{Acknowledgements}

The ALICE Collaboration would like to thank all its engineers and technicians for their invaluable contributions to the construction of the experiment and the CERN accelerator teams for the outstanding performance of the LHC complex.
The ALICE Collaboration gratefully acknowledges the resources and support provided by all Grid centres and the Worldwide LHC Computing Grid (WLCG) collaboration.
The ALICE Collaboration acknowledges the following funding agencies for their support in building and running the ALICE detector:
A. I. Alikhanyan National Science Laboratory (Yerevan Physics Institute) Foundation (ANSL), State Committee of Science and World Federation of Scientists (WFS), Armenia;
Austrian Academy of Sciences and Nationalstiftung f\"{u}r Forschung, Technologie und Entwicklung, Austria;
Ministry of Communications and High Technologies, National Nuclear Research Center, Azerbaijan;
Conselho Nacional de Desenvolvimento Cient\'{\i}fico e Tecnol\'{o}gico (CNPq), Universidade Federal do Rio Grande do Sul (UFRGS), Financiadora de Estudos e Projetos (Finep) and Funda\c{c}\~{a}o de Amparo \`{a} Pesquisa do Estado de S\~{a}o Paulo (FAPESP), Brazil;
Ministry of Science \& Technology of China (MSTC), National Natural Science Foundation of China (NSFC) and Ministry of Education of China (MOEC) , China;
Ministry of Science, Education and Sport and Croatian Science Foundation, Croatia;
Ministry of Education, Youth and Sports of the Czech Republic, Czech Republic;
The Danish Council for Independent Research | Natural Sciences, the Carlsberg Foundation and Danish National Research Foundation (DNRF), Denmark;
Helsinki Institute of Physics (HIP), Finland;
Commissariat \`{a} l'Energie Atomique (CEA) and Institut National de Physique Nucl\'{e}aire et de Physique des Particules (IN2P3) and Centre National de la Recherche Scientifique (CNRS), France;
Bundesministerium f\"{u}r Bildung, Wissenschaft, Forschung und Technologie (BMBF) and GSI Helmholtzzentrum f\"{u}r Schwerionenforschung GmbH, Germany;
Ministry of Education, Research and Religious Affairs, Greece;
National Research, Development and Innovation Office, Hungary;
Department of Atomic Energy Government of India (DAE) and Council of Scientific and Industrial Research (CSIR), New Delhi, India;
Indonesian Institute of Science, Indonesia;
Centro Fermi - Museo Storico della Fisica e Centro Studi e Ricerche Enrico Fermi and Istituto Nazionale di Fisica Nucleare (INFN), Italy;
Institute for Innovative Science and Technology , Nagasaki Institute of Applied Science (IIST), Japan Society for the Promotion of Science (JSPS) KAKENHI and Japanese Ministry of Education, Culture, Sports, Science and Technology (MEXT), Japan;
Consejo Nacional de Ciencia (CONACYT) y Tecnolog\'{i}a, through Fondo de Cooperaci\'{o}n Internacional en Ciencia y Tecnolog\'{i}a (FONCICYT) and Direcci\'{o}n General de Asuntos del Personal Academico (DGAPA), Mexico;
Nationaal instituut voor subatomaire fysica (Nikhef), Netherlands;
The Research Council of Norway, Norway;
Commission on Science and Technology for Sustainable Development in the South (COMSATS), Pakistan;
Pontificia Universidad Cat\'{o}lica del Per\'{u}, Peru;
Ministry of Science and Higher Education and National Science Centre, Poland;
Korea Institute of Science and Technology Information and National Research Foundation of Korea (NRF), Republic of Korea;
Ministry of Education and Scientific Research, Institute of Atomic Physics and Romanian National Agency for Science, Technology and Innovation, Romania;
Joint Institute for Nuclear Research (JINR), Ministry of Education and Science of the Russian Federation and National Research Centre Kurchatov Institute, Russia;
Ministry of Education, Science, Research and Sport of the Slovak Republic, Slovakia;
National Research Foundation of South Africa, South Africa;
Centro de Aplicaciones Tecnol\'{o}gicas y Desarrollo Nuclear (CEADEN), Cubaenerg\'{\i}a, Cuba, Ministerio de Ciencia e Innovacion and Centro de Investigaciones Energ\'{e}ticas, Medioambientales y Tecnol\'{o}gicas (CIEMAT), Spain;
Swedish Research Council (VR) and Knut \& Alice Wallenberg Foundation (KAW), Sweden;
European Organization for Nuclear Research, Switzerland;
National Science and Technology Development Agency (NSDTA), Suranaree University of Technology (SUT) and Office of the Higher Education Commission under NRU project of Thailand, Thailand;
Turkish Atomic Energy Agency (TAEK), Turkey;
National Academy of  Sciences of Ukraine, Ukraine;
Science and Technology Facilities Council (STFC), United Kingdom;
National Science Foundation of the United States of America (NSF) and United States Department of Energy, Office of Nuclear Physics (DOE NP), United States of America.
\end{acknowledgement}

\bibliography{mybib9}{}
\bibliographystyle{utphys}

%
%
%
\newpage
\appendix
\section{The ALICE Collaboration}
\label{app:collab}



\begingroup
\small
\begin{flushleft}
D.~Adamov\'{a}$^\textrm{\scriptsize 87}$,
M.M.~Aggarwal$^\textrm{\scriptsize 91}$,
G.~Aglieri Rinella$^\textrm{\scriptsize 34}$,
M.~Agnello$^\textrm{\scriptsize 30}$\textsuperscript{,}$^\textrm{\scriptsize 113}$,
N.~Agrawal$^\textrm{\scriptsize 47}$,
Z.~Ahammed$^\textrm{\scriptsize 139}$,
S.~Ahmad$^\textrm{\scriptsize 17}$,
S.U.~Ahn$^\textrm{\scriptsize 69}$,
S.~Aiola$^\textrm{\scriptsize 143}$,
A.~Akindinov$^\textrm{\scriptsize 54}$
% ...
\renewcommand\labelenumi{\textsuperscript{\theenumi}~}

\section*{Affiliation notes}
\renewcommand\theenumi{\roman{enumi}}
\begin{Authlist}
\item \Adef{0}Deceased
\item \Adef{idp1804352}{Also at: Georgia State University, Atlanta, Georgia, United States}
\item \Adef{idp3241008}{Also at: Also at Department of Applied Physics, Aligarh Muslim University, Aligarh, India}
\item \Adef{idp4012544}{Also at: M.V. Lomonosov Moscow State University, D.V. Skobeltsyn Institute of Nuclear, Physics, Moscow, Russia}
\end{Authlist}

\section*{Collaboration Institutes}
\renewcommand\theenumi{\arabic{enumi}~}

$^{1}$A.I. Alikhanyan National Science Laboratory (Yerevan Physics Institute) Foundation, Yerevan, Armenia
\\
$^{2}$Benem\'{e}rita Universidad Aut\'{o}noma de Puebla, Puebla, Mexico
\\
$^{3}$Bogolyubov Institute for Theoretical Physics, Kiev, Ukraine
\\
$^{4}$Bose Institute, Department of Physics 
and Centre for Astroparticle Physics and Space Science (CAPSS), Kolkata, India
\\
$^{5}$Budker Institute for Nuclear Physics, Novosibirsk, Russia
\\
$^{6}$California Polytechnic State University, San Luis Obispo, California, United States
\\
$^{7}$Central China Normal University, Wuhan, China
\\
$^{8}$Centre de Calcul de l'IN2P3, Villeurbanne, Lyon, France
\\
$^{9}$Centro de Aplicaciones Tecnol\'{o}gicas y Desarrollo Nuclear (CEADEN), Havana, Cuba
\\
$^{10}$Centro de Investigaciones Energ\'{e}ticas Medioambientales y Tecnol\'{o}gicas (CIEMAT), Madrid, Spain
\\
$^{11}$Centro de Investigaci\'{o}n y de Estudios Avanzados (CINVESTAV), Mexico City and M\'{e}rida, Mexico
\\
$^{12}$Centro Fermi - Museo Storico della Fisica e Centro Studi e Ricerche ``Enrico Fermi', Rome, Italy
\\
$^{13}$Chicago State University, Chicago, Illinois, United States
\\
$^{14}$China Institute of Atomic Energy, Beijing, China
\\
$^{15}$COMSATS Institute of Information Technology (CIIT), Islamabad, Pakistan
\\
$^{16}$Departamento de F\'{\i}sica de Part\'{\i}culas and IGFAE, Universidad de Santiago de Compostela, Santiago de Compostela, Spain
\\
$^{17}$Department of Physics, Aligarh Muslim University, Aligarh, India
\\
$^{18}$Department of Physics, Ohio State University, Columbus, Ohio, United States
\\
$^{19}$Department of Physics, Sejong University, Seoul, South Korea
\\
$^{20}$Department of Physics, University of Oslo, Oslo, Norway
\\
$^{21}$Department of Physics and Technology, University of Bergen, Bergen, Norway
\\
$^{22}$Dipartimento di Fisica dell'Universit\`{a} 'La Sapienza'
and Sezione INFN, Rome, Italy
\\
$^{23}$Dipartimento di Fisica dell'Universit\`{a}
and Sezione INFN, Cagliari, Italy
\\
$^{24}$Dipartimento di Fisica dell'Universit\`{a}
and Sezione INFN, Trieste, Italy
\\
$^{25}$Dipartimento di Fisica dell'Universit\`{a}
and Sezione INFN, Turin, Italy
\\
$^{26}$Dipartimento di Fisica e Astronomia dell'Universit\`{a}
and Sezione INFN, Bologna, Italy
\\
$^{27}$Dipartimento di Fisica e Astronomia dell'Universit\`{a}
and Sezione INFN, Catania, Italy
\\
$^{28}$Dipartimento di Fisica e Astronomia dell'Universit\`{a}
and Sezione INFN, Padova, Italy
\\
$^{29}$Dipartimento di Fisica `E.R.~Caianiello' dell'Universit\`{a}
and Gruppo Collegato INFN, Salerno, Italy
\\
$^{30}$Dipartimento DISAT del Politecnico and Sezione INFN, Turin, Italy
\\
$^{31}$Dipartimento di Scienze e Innovazione Tecnologica dell'Universit\`{a} del Piemonte Orientale and INFN Sezione di Torino, Alessandria, Italy
\\
$^{32}$Dipartimento Interateneo di Fisica `M.~Merlin'
and Sezione INFN, Bari, Italy
\\
$^{33}$Division of Experimental High Energy Physics, University of Lund, Lund, Sweden
\\
$^{34}$European Organization for Nuclear Research (CERN), Geneva, Switzerland
\\
$^{35}$Excellence Cluster Universe, Technische Universit\"{a}t M\"{u}nchen, Munich, Germany
\\
$^{36}$Faculty of Engineering, Bergen University College, Bergen, Norway
\\
$^{37}$Faculty of Mathematics, Physics and Informatics, Comenius University, Bratislava, Slovakia
\\
$^{38}$Faculty of Nuclear Sciences and Physical Engineering, Czech Technical University in Prague, Prague, Czech Republic
\\
$^{39}$Faculty of Science, P.J.~\v{S}af\'{a}rik University, Ko\v{s}ice, Slovakia
\\
$^{40}$Faculty of Technology, Buskerud and Vestfold University College, Tonsberg, Norway
\\
$^{41}$Frankfurt Institute for Advanced Studies, Johann Wolfgang Goethe-Universit\"{a}t Frankfurt, Frankfurt, Germany
\\
$^{42}$Gangneung-Wonju National University, Gangneung, South Korea
\\
$^{43}$Gauhati University, Department of Physics, Guwahati, India
\\
$^{44}$Helmholtz-Institut f\"{u}r Strahlen- und Kernphysik, Rheinische Friedrich-Wilhelms-Universit\"{a}t Bonn, Bonn, Germany
\\
$^{45}$Helsinki Institute of Physics (HIP), Helsinki, Finland
\\
$^{46}$Hiroshima University, Hiroshima, Japan
\\
$^{47}$Indian Institute of Technology Bombay (IIT), Mumbai, India
\\
$^{48}$Indian Institute of Technology Indore, Indore, India
\\
$^{49}$Indonesian Institute of Sciences, Jakarta, Indonesia
\\
$^{50}$Inha University, Incheon, South Korea
\\
$^{51}$Institut de Physique Nucl\'eaire d'Orsay (IPNO), Universit\'e Paris-Sud, CNRS-IN2P3, Orsay, France
\\
$^{52}$Institute for Nuclear Research, Academy of Sciences, Moscow, Russia
\\
$^{53}$Institute for Subatomic Physics of Utrecht University, Utrecht, Netherlands
\\
$^{54}$Institute for Theoretical and Experimental Physics, Moscow, Russia
\\
$^{55}$Institute of Experimental Physics, Slovak Academy of Sciences, Ko\v{s}ice, Slovakia
\\
$^{56}$Institute of Physics, Academy of Sciences of the Czech Republic, Prague, Czech Republic
\\
$^{57}$Institute of Physics, Bhubaneswar, India
\\
$^{58}$Institute of Space Science (ISS), Bucharest, Romania
\\
$^{59}$Institut f\"{u}r Informatik, Johann Wolfgang Goethe-Universit\"{a}t Frankfurt, Frankfurt, Germany
\\
$^{60}$Institut f\"{u}r Kernphysik, Johann Wolfgang Goethe-Universit\"{a}t Frankfurt, Frankfurt, Germany
\\
$^{61}$Institut f\"{u}r Kernphysik, Westf\"{a}lische Wilhelms-Universit\"{a}t M\"{u}nster, M\"{u}nster, Germany
\\
$^{62}$Instituto de Ciencias Nucleares, Universidad Nacional Aut\'{o}noma de M\'{e}xico, Mexico City, Mexico
\\
$^{63}$Instituto de F\'{i}sica, Universidade Federal do Rio Grande do Sul (UFRGS), Porto Alegre, Brazil
\\
$^{64}$Instituto de F\'{\i}sica, Universidad Nacional Aut\'{o}noma de M\'{e}xico, Mexico City, Mexico
\\
$^{65}$IRFU, CEA, Universit\'{e} Paris-Saclay, F-91191 Gif-sur-Yvette, France, Saclay, France
\\
$^{66}$iThemba LABS, National Research Foundation, Somerset West, South Africa
\\
$^{67}$Joint Institute for Nuclear Research (JINR), Dubna, Russia
\\
$^{68}$Konkuk University, Seoul, South Korea
\\
$^{69}$Korea Institute of Science and Technology Information, Daejeon, South Korea
\\
$^{70}$KTO Karatay University, Konya, Turkey
\\
$^{71}$Laboratoire de Physique Corpusculaire (LPC), Clermont Universit\'{e}, Universit\'{e} Blaise Pascal, CNRS--IN2P3, Clermont-Ferrand, France
\\
$^{72}$Laboratoire de Physique Subatomique et de Cosmologie, Universit\'{e} Grenoble-Alpes, CNRS-IN2P3, Grenoble, France
\\
$^{73}$Laboratori Nazionali di Frascati, INFN, Frascati, Italy
\\
$^{74}$Laboratori Nazionali di Legnaro, INFN, Legnaro, Italy
\\
$^{75}$Lawrence Berkeley National Laboratory, Berkeley, California, United States
\\
$^{76}$Moscow Engineering Physics Institute, Moscow, Russia
\\
$^{77}$Nagasaki Institute of Applied Science, Nagasaki, Japan
\\
$^{78}$National and Kapodistrian University of Athens, Physics Department, Athens, Greece, Athens, Greece
\\
$^{79}$National Centre for Nuclear Studies, Warsaw, Poland
\\
$^{80}$National Institute for Physics and Nuclear Engineering, Bucharest, Romania
\\
$^{81}$National Institute of Science Education and Research, Bhubaneswar, India
\\
$^{82}$National Nuclear Research Center, Baku, Azerbaijan
\\
$^{83}$National Research Centre Kurchatov Institute, Moscow, Russia
\\
$^{84}$Niels Bohr Institute, University of Copenhagen, Copenhagen, Denmark
\\
$^{85}$Nikhef, Nationaal instituut voor subatomaire fysica, Amsterdam, Netherlands
\\
$^{86}$Nuclear Physics Group, STFC Daresbury Laboratory, Daresbury, United Kingdom
\\
$^{87}$Nuclear Physics Institute, Academy of Sciences of the Czech Republic, \v{R}e\v{z} u Prahy, Czech Republic
\\
$^{88}$Oak Ridge National Laboratory, Oak Ridge, Tennessee, United States
\\
$^{89}$Petersburg Nuclear Physics Institute, Gatchina, Russia
\\
$^{90}$Physics Department, Creighton University, Omaha, Nebraska, United States
\\
$^{91}$Physics Department, Panjab University, Chandigarh, India
\\
$^{92}$Physics Department, University of Cape Town, Cape Town, South Africa
\\
$^{93}$Physics Department, University of Jammu, Jammu, India
\\
$^{94}$Physics Department, University of Rajasthan, Jaipur, India
\\
$^{95}$Physikalisches Institut, Eberhard Karls Universit\"{a}t T\"{u}bingen, T\"{u}bingen, Germany
\\
$^{96}$Physikalisches Institut, Ruprecht-Karls-Universit\"{a}t Heidelberg, Heidelberg, Germany
\\
$^{97}$Physik Department, Technische Universit\"{a}t M\"{u}nchen, Munich, Germany
\\
$^{98}$Purdue University, West Lafayette, Indiana, United States
\\
$^{99}$Pusan National University, Pusan, South Korea
\\
$^{100}$Research Division and ExtreMe Matter Institute EMMI, GSI Helmholtzzentrum f\"ur Schwerionenforschung GmbH, Darmstadt, Germany
\\
$^{101}$Rudjer Bo\v{s}kovi\'{c} Institute, Zagreb, Croatia
\\
$^{102}$Russian Federal Nuclear Center (VNIIEF), Sarov, Russia
\\
$^{103}$Saha Institute of Nuclear Physics, Kolkata, India
\\
$^{104}$School of Physics and Astronomy, University of Birmingham, Birmingham, United Kingdom
\\
$^{105}$Secci\'{o}n F\'{\i}sica, Departamento de Ciencias, Pontificia Universidad Cat\'{o}lica del Per\'{u}, Lima, Peru
\\
$^{106}$Sezione INFN, Bari, Italy
\\
$^{107}$Sezione INFN, Bologna, Italy
\\
$^{108}$Sezione INFN, Cagliari, Italy
\\
$^{109}$Sezione INFN, Catania, Italy
\\
$^{110}$Sezione INFN, Padova, Italy
\\
$^{111}$Sezione INFN, Rome, Italy
\\
$^{112}$Sezione INFN, Trieste, Italy
\\
$^{113}$Sezione INFN, Turin, Italy
\\
$^{114}$SSC IHEP of NRC Kurchatov institute, Protvino, Russia
\\
$^{115}$Stefan Meyer Institut f\"{u}r Subatomare Physik (SMI), Vienna, Austria
\\
$^{116}$SUBATECH, Ecole des Mines de Nantes, Universit\'{e} de Nantes, CNRS-IN2P3, Nantes, France
\\
$^{117}$Suranaree University of Technology, Nakhon Ratchasima, Thailand
\\
$^{118}$Technical University of Ko\v{s}ice, Ko\v{s}ice, Slovakia
\\
$^{119}$Technical University of Split FESB, Split, Croatia
\\
$^{120}$The Henryk Niewodniczanski Institute of Nuclear Physics, Polish Academy of Sciences, Cracow, Poland
\\
$^{121}$The University of Texas at Austin, Physics Department, Austin, Texas, United States
\\
$^{122}$Universidad Aut\'{o}noma de Sinaloa, Culiac\'{a}n, Mexico
\\
$^{123}$Universidade de S\~{a}o Paulo (USP), S\~{a}o Paulo, Brazil
\\
$^{124}$Universidade Estadual de Campinas (UNICAMP), Campinas, Brazil
\\
$^{125}$Universidade Federal do ABC, Santo Andre, Brazil
\\
$^{126}$University of Houston, Houston, Texas, United States
\\
$^{127}$University of Jyv\"{a}skyl\"{a}, Jyv\"{a}skyl\"{a}, Finland
\\
$^{128}$University of Liverpool, Liverpool, United Kingdom
\\
$^{129}$University of Tennessee, Knoxville, Tennessee, United States
\\
$^{130}$University of the Witwatersrand, Johannesburg, South Africa
\\
$^{131}$University of Tokyo, Tokyo, Japan
\\
$^{132}$University of Tsukuba, Tsukuba, Japan
\\
$^{133}$University of Zagreb, Zagreb, Croatia
\\
$^{134}$Universit\'{e} de Lyon, Universit\'{e} Lyon 1, CNRS/IN2P3, IPN-Lyon, Villeurbanne, Lyon, France
\\
$^{135}$Universit\'{e} de Strasbourg, CNRS, IPHC UMR 7178, F-67000 Strasbourg, France, Strasbourg, France
\\
$^{136}$Universit\`{a} degli Studi di Pavia, Pavia, Italy
\\
$^{137}$Universit\`{a} di Brescia, Brescia, Italy
\\
$^{138}$V.~Fock Institute for Physics, St. Petersburg State University, St. Petersburg, Russia
\\
$^{139}$Variable Energy Cyclotron Centre, Kolkata, India
\\
$^{140}$Warsaw University of Technology, Warsaw, Poland
\\
$^{141}$Wayne State University, Detroit, Michigan, United States
\\
$^{142}$Wigner Research Centre for Physics, Hungarian Academy of Sciences, Budapest, Hungary
\\
$^{143}$Yale University, New Haven, Connecticut, United States
\\
$^{144}$Yonsei University, Seoul, South Korea
\\
$^{145}$Zentrum f\"{u}r Technologietransfer und Telekommunikation (ZTT), Fachhochschule Worms, Worms, Germany
\endgroup

\endgroup

\end{document}